# Transition from Eyeball to Snowball Driven by Sea-ice Drift on Tidally Locked Terrestrial Planets

Jun Yang[1,*], Weiwen Ji[1], & Yaoxuan Zeng[1]

[1]Department of Atmospheric and Oceanic Sciences, School of Physics, Peking University, Beijing, 100871, China.

[*]Corresponding author: J.Y., junyang@pku.edu.cn

**Tidally locked terrestrial planets around low-mass stars are the prime targets for future atmospheric characterizations of potentially habitable systems[1], especially the three nearby ones–Proxima b[2], TRAPPIST-1e[3], and LHS 1140b[4]. Previous studies suggest that if these planets have surface ocean they would be in an eyeball-like climate state[5-10]: ice-free in the vicinity of the substellar point and ice-covered in the rest regions. However, an important component of the climate system–sea ice dynamics has not been well studied in previous studies. A fundamental question is: would the open ocean be stable against a globally ice-covered snowball state? Here we show that sea-ice drift cools the ocean's surface when the ice flows to the warmer substellar region and melts through absorbing heat from the ocean and the overlying air. As a result, the open ocean shrinks and can even disappear when atmospheric greenhouse gases are not much more abundant than on Earth, turning the planet into a snowball state. This occurs for both synchronous rotation and spin-orbit resonances (such as 3:2). These results suggest that sea-ice drift strongly reduces the open ocean area and can significantly impact the habitability of tidally locked planets.**

Sea-ice drift, driven by surface winds and ocean currents, transports heat and freshwater across the ocean surface, directly or indirectly influencing ice concentration, ice growth and melt, ice thickness, surface albedo, and air–sea heat exchange[11,12]. As a result, sea-ice drift plays critical roles in the Earth climate system, such as a snowball Earth possibly occurred in 630–750 million years ago[13,14]. When sea-ice drift is considered, the $CO_2$ concentration threshold for the snowball Earth formation is ≈100 times higher, compared to the simulations without sea-ice drift[15,16]. In this study, we examine the effect of sea-ice drift on tidally locked planets around M dwarfs. We perform three-dimensional (3D) global climate modeling of nine rocky planets orbiting in the outer region of their stars' habitable zones in synchronous rotation or in 3:2 spin-orbit resonance (Supplementary Table 1). The effect of sea-ice drift is explored through comparing the experiments with sea-ice drift to those without sea-ice drift (see Methods).

In the 3D atmosphere-only climate experiments (the roles of ocean and sea ice dynamics are excluded), the planets have an ocean of open water under the substellar point while the rest regions are ice-covered, called an eyeball climate state[5-9] (Fig. 1 and Supplementary Figs. 1 & 2). When ocean dynamics are also included, the global-mean ice coverage generally decreases and the spatial pattern of the open ocean changes (similar to a 'lobster' in some cases), due to that oceanic currents and Rossby and Kelvin waves transport heat from the substellar region to the ice margins and melt the ice there[10,17]. When sea ice dynamics are further included, TRAPPIST-1e, Kepler-1229b, LHS 1140b, Kapteyn b, and TRAPPIST-1f enter a snowball state with ice coverage of ≈100%, and the ice coverage of Proxima b and Wolf 1061c increases from 69% to 72% and 67% to 75%, respectively. These results indicate that sea-ice drift acts to shrink the open ocean whereas oceanic heat transport acts to expand the open ocean. The effect of the former is stronger than the latter for the planets those receive relatively lower stellar fluxes (less than ≈800 W m$^{-2}$). Kepler-442b and Proxima b do not enter a snowball state mainly because of their high stellar fluxes, 956 and 887 W m$^{-2}$, respectively.

The importance of sea-ice drift is further confirmed in examining the stellar flux threshold for the onset of a snowball glaciation. When Earth's radius and gravity and a rotation period of 10 Earth days are employed, the stellar flux threshold is 500 W m$^{-2}$ in both atmosphere-only experiments and coupled atmosphere–ocean experiments, but it is 800 W m$^{-2}$ in the fully coupled atmosphere–ocean–sea-ice experiments (Supplementary Fig. 3). This indicates that sea-ice drift has the equivalent effect of a stellar radiation of −300 W m$^{-2}$ in driving the system into a snowball state.

Sensitivity tests show that the drift of sea ice is efficient in increasing the ice coverage for a wide range of parameters, including planetary rotation period, gravity, and radius (Supplementary Figs. 4 to 7). In the fully coupled atmosphere–ocean–sea-ice experiments sea-ice drift increases the global-mean ice coverage by ≈5–30%, compared to the coupled atmosphere–ocean experiments without sea ice dynamics. In both regimes of atmospheric circulation, rapidly rotating ($L_R < R_p$, where $L_R$ is the equatorial Rossby deformation radius[18,19] and $R_p$ is planetary radius) and slowly rotating ($L_R > R_p$), sea-ice drift would shrink the open ocean area. Moreover, the results do not depend on internal ice compressive strength (Supplementary Fig. 8d), which is the main tuning parameter in the module of sea ice dynamics[12]. This is due to the fact that the ice is $\mathcal{O}(1\text{–}10 \text{ m})$ thick (Supplementary Fig. 9)[10,17] and the primary forces responsible for the ice motion are air–ice stress and ocean–ice stress while the internal ice stress is much weaker[20].

Why are the sea-ice dynamics so efficient to lead the system to a snowball state? The answer is the combined effect of heat absorption during sea ice melting and surface albedo increasing associated with sea-ice flows (Fig. 2). For a synchronous rotation orbit, the nightside is much colder than the dayside, so that the horizontal surface temperature gradients are large especially around the terminators (Supplementary Fig. 2). The strong temperature gradients drive robust surface winds that transport ice from the colder regions where the ice forms to the warmer regions around the substellar point (Fig. 2f). Part of the drifted ice persists and leads to increasing surface albedo. Part of the drifted ice is melted and leads to cooling the ocean (Fig. 2b) because the melting process absorbs heat from the ocean and the overlying atmosphere, and then new drifting ice flows to the melted areas. The heat uptake near the ice edges is as large as $\mathcal{O}(10\text{-}100\ \mathrm{W\ m^{-2}})$ in the experiments with sea-ice drift whereas it is one order smaller in the runs without sea-ice drift. This cooling effect is more effective than the ice albedo feedback (Fig. 2c), which is weak due to the low ice albedo ($\approx$0.1–0.35) under M-dwarf spectra[21] and meanwhile the masking effect of clouds. Optically thick clouds form on the dayside[22], which reduces the contribution of the surface to planetary albedo. This masking effect is further confirmed in the experiments of varying the ice albedo (Supplementary Fig. 10). Increasing the ice albedo has no significant effects on the climate as long as the surface albedo is lower than atmospheric reflection associated with clouds and Rayleigh scattering.

Continents can influence the open ocean area. For example, an open ocean could persist at the substellar region on TRAPPIST-1e when continents are included in the simulations (Fig. 3). The open ocean area is smaller when the substellar point is located over land than over ocean. In particular, when a super-continent covers the substellar region, the planet enters a snowball state (Supplementary Fig. 11). This implies that tidally locked planets are more ready to enter a snowball state when plate tectonics drive the continent(s) to the dayside. In these experiments, it is still clear to see that sea-ice drift acts to increase the ice coverage, comparing to the experiments without sea-ice drift. The robustness of the conclusion is due to the simplicity of the mechanism: at the ice edges the direction of surface winds is always from cold regions (corresponding to high pressures) to warm regions (low pressures)[23], and these winds drive the sea ice flowing toward the substellar region. The effect of sea-ice drift, as well as oceanic heat transport, however, is weaker than that in the aqua-planet configuration, due to the blocking and friction of continents.

For a spin-orbit resonance orbit, such as 3:2 like Mercury, all longitudes receive stellar radiation but time mean instellation at the equator is much lower than that at the permanent substellar point of a 1:1 orbit. In the experiments without sea-ice drift, Proxima b and TRAPPIST-1e exhibit a climate with tropical waterbelts[6,8-10] due to their short solar days, ≈22 and 12 Earth days, but Kepler-442b and Wolf 1061c have elliptic open oceans those move following the substellar point because of their longer solar days, 224 and 36 Earth days (Fig. 4). Sea-ice drift, however, drives TRAPPIST-1e and Kepler-442b into a snowball state and increases the ice coverage of Proxima b and Wolf 1061c from 48% to 56% and 67% to 69%, respectively. LHS 1140b, as well as Kepler-1229b, Kapteyn b, TRAPPIST-1f, and Kepler-186f, enters a snowball state regardless with or without sea-ice drift due to its low stellar flux.

For both 1:1 and 3:2 orbits, one planet having a smaller (larger) planetary radius would have a higher (lower) sea ice coverage (Supplementary Figs. 4, 7, & 16). The main reason is that varying radius influences atmospheric and oceanic circulations and oceanic heat transport becomes stronger with increasing radius (Supplementary Figs. 5 & 6). This is why TRAPPIST-1e enters a snowball state but not for Wolf 1061c (Figs. 1 & 4), although they have nearly the same stellar flux, 821 and 819 W m$^{-2}$, respectively. TRAPPIST-1e's radius is only 56% of Wolf 1061c. Moreover, varying rotation period influences the spatial pattern of sea ice but does not significantly affect the global-mean ice coverage under 1:1 orbit (Supplementary Figs. 4 & 5), however, the global-mean ice coverage greatly increases with rotation period in the 3:2 orbit experiments (Supplementary Figs. 14 & 15). Rotation period determines the length of day and night ($L_{day} = P_{orb}P_{rot}/(P_{orb} - P_{rot})$ where $P_{rot}$ is the rotation period and $P_{orb}$ is the orbital period) for a resonance orbit, but not for a synchronous orbit. For a 3:2 orbit, $L_{day}$ is three times $P_{rot}$. More ice could grow during a longer night and the ice reflects more stellar radiation back to space during the day, especially during the morning time[24]. This is one reason for that Kepler-442b has much higher ice coverage than Proxima b (Fig. 4), although Kepler-442b's stellar flux is higher. Their day lengths are 224 and 22 Earth days, respectively. The other reason is that ice albedos on Kepler-442b are higher, (0.75, 0.30) versus (0.65, 0.15). The ice albedo effect in the 3:2 orbit is stronger than that in the 1:1 orbit, because more stellar radiation can be deposited on the ice when the substellar point keeps moving rather than stays still[24].

We further find that a high background air mass (such as $N_2$) could also drive the system to a snowball state. For example, the ice coverage on LHS 1140b increases with background air mass and even reaches 100% when the air mass is ≥3 times Earth's value (Supplementary Fig. 3d). The main reason is that increasing air mass enhances planetary albedo through Rayleigh scattering and raises atmospheric heat transport[25] from the dayside to the nightside, warming the nightside surface but cooling the dayside surface (Supplementary Fig. 18). Therefore, the sea ice is relatively easier to expand toward the substellar region when the ice edges have already crossed the terminators, promoting the snowball formation. The warming effect of pressure broadening[26] is weak because water vapor concentration is low in these experiments.

Due to the effect of sea ice dynamics, greenhouse gas concentration required to maintain a significantly large open-ocean area is higher (or volcanic outgassing rate should be faster if an active carbon cycle[27,28] is considered) than that in the experiments without sea-ice drift. For example, $CO_2$ mixing ratio in the experiments with sea-ice drift should be increased from 300 to 19,200 ppmv in order to maintain an open ocean of ≈20% of the global area on TRAPPIST-1e (Supplementary Fig. 8 vs Fig. 1). For LHS 1140b, the $CO_2$ mixing ratio has to be increased from 300 to 76,800 ppmv in addition to $CH_4$ increased from 0.8 to 1000 ppmv to have an open ocean of ≈15% of the global area.

How to distinguish the three climates, eyeball, lobster, and snowball, in observations? In the snowball state, water vapor concentration is lower and day–night thermal emission contrast is smaller than those in the other two states (Supplementary Fig. 19). Unfortunately, the magnitude of the differences is small and differentiating them may require the *James Webb Space Telescope* or ground-based extreme large telescopes[1,29]. In planetary albedo (for reflectance phase curve observations), the differences are also small because of the masking effect of clouds mentioned above. It is interesting to note that the peak of the thermal phase curve of Proxima b exhibits a negative phase angle displacement of ≈30º in the coupled atmosphere–ocean experiment but a positive displacement of ≈45º in the fully coupled atmosphere–ocean–sea-ice experiment. The former is mainly due to that tropical westerly winds transport clouds to the east side of the substellar point, where the clouds emit thermal radiation to space at low temperatures[19,22]. The latter is because sea-ice drift significantly increases the ice concentrations at the west side of the substellar point (Fig. 1), cooling the surface there (Supplementary Fig. 2) and allowing less thermal emission to space.

Finally, we note that a snowball state is an extreme stressor on life and their evolution but it does not imply the system is uninhabitable completely. Photosynthesis organism may develop in thin ice regions where stellar radiation could penetrate the ice (such as in the vicinity of the substellar point), at local ice-free regions where there is a high-level geothermal heat, or at constricted marginal seas where ice drift is limited by the sidewalls of the sea[30]. Future work is required to estimate the effect of ice sheet dynamics.

**Acknowledgements** We are grateful to Feng Ding, Thomas J. Fauchez, Yonggang Liu, and Jintai Lin for fruitful discussions with them, to Yongyun Hu and Yosef Ashkenazy for their great help in improving the manuscript, and to Yuwei Wang for his help in modifying source codes of the model. J.Y. acknowledges supports from the National Natural Science Foundation of China (NSFC) grants 41861124002, 41675071, 41606060, and 41761144072.

**Author Contributions** J.Y. formulated the problem, designed the experiments, analyzed the data, and wrote the manuscript. J.Y. and W.J. performed the numerical experiments. All authors contributed to data analyses.

**Materials and Correspondence** should be addressed to J.Y., junyang@pku.edu.cn.

**Competing Interests** The authors declare that they have no competing financial interests.

**Figure Captions:**

**Fig. 1**. **Sea ice concentrations in synchronous rotation orbits.** Left column: the results of 3D atmosphere-only simulations (denoted by 'Atm'); center column: coupled atmosphere-ocean simulations ('Atm + Ocn'); right column: fully coupled atmosphere-ocean-sea-ice simulations ('Atm + Ocn + Ice Drift'). From top to bottom, the planets are Kepler-442b, Proxima b, TRAPPIST-1e, Wolf 1061c, Kepler-1229b, LHS 1140b, Kapteyn b$^*$ (unconfirmed), TRAPPIST-1f, and Kepler-186f, respectively. They are listed in the order of decreasing stellar flux (Supplementary Table 1). In each panel, the x-axis is longitude from 0° to 360°, the y-axis is latitude from 90°S to 90°N, and the red dot is the substellar point. The number in the upper left corner of each panel is the mean ice coverage over the ocean. By default, the column air mass is $1.0 \times 10^4$ kg m$^{-2}$ (equal to modern Earth), column $CO_2$ mass is ≈4.5 kg m$^{-2}$ (corresponding to a mixing ratio of 300 ppmv), column $CH_4$ mass is ≈0.004 kg m$^{-2}$ (0.8 ppmv), the surface is a water world with an ocean depth of ≈1000 m (uniform), and rotation period is equal to orbital period. For TRAPPIST-1f and Kepler-186f, the column $CO_2$ mass is set to ≈1500 kg m$^{-2}$ (or $10^5$ ppmv) and the column $CH_4$ mass ≈5 kg m$^{-2}$ (or 1000 ppmv) in order to compensate their low stellar fluxes. Ocean dynamics expand the open ocean area but sea ice dynamics reduce the open ocean area. Exceptionally, in the 'Atm + Ocn + Ice Drift' run of Kepler-442b, the sea ice coverage is 2% lower than that of the 'Atm + Ocn' run, due to that ocean currents are relatively stronger in the former experiment. For the details of how planetary radius, rotation period, and gravity influence the results, please the Supplementary Figs. 4 to 7.

**Fig. 2**. **Physical mechanisms for the effect of sea-ice drift. a-d**, time-series of global-mean sea ice coverage (**a**), area-averaged surface temperature in the vicinity of the substellar point from 30°S to 30°N and 150° to 210° in longitude (**b**), global-mean surface albedo (**c**), and planetary albedo (**d**) in the experiments of LHS 1140b. **e-f**, spatial pattern of heat release during sea ice formation (positive values) and heat uptake during sea ice melting (negative values) in the 5$^{th}$ yr of the experiments. **e** and the red lines in **a-d** are for the coupled atmosphere-ocean experiments, and **f** and the blue lines in **a-d** are for the fully coupled atmosphere-ocean-sea-ice experiments. The vectors in **f** denote sea ice velocities. The thin lines in **e-f** are the ice margins with an ice concentration of 50%, and the red dot is the substellar point. Note that in **d**, the planetary albedo is relatively lower in the snowball state (blue line) because of the reduction of cloud water path. Sea ice dynamics increase the ice coverage through increasing the surface albedo and cooling the sea surface during ice melting. For the entire evolution process, please watch the Supplementary Video 1 online.

**Fig. 3**. **Effects of continents on the sea ice concentrations**. 'Atm + Ocn': coupled atmosphere-ocean experiments without sea-ice drift, and 'Atm + Ocn + Ice Drift': fully coupled atmosphere-ocean-sea-ice experiments with sea-ice drift. Two different configurations (gray color) are used: modern Earth and 630 Ma Earth. The red dot is the substellar point, which locates near the Africa, at the center of the Atlantic Ocean, at the center of an open ocean, and over one continent, respectively. The vectors denote sea ice velocities. The number in the upper left corner of each panel is the mean ice coverage over the ocean. The parameters of TRAPPIST-1e are employed, column air mass is 1.0 × 10$^4$ kg m$^{-2}$, and column $CO_2$ mass is ≈4.5 kg m$^{-2}$ in all these experiments. Realistic ocean depths are used in the upper two panels, and a uniform ocean depth of 4000 m is used in the lower two panels. Tests with idealized continents show the same trend in the effect of sea-ice drift (Supplementary Fig. 11). A relatively low resolution was employed (see Methods); increasing the model resolution has a very small effect on the results, as shown in Supplementary Fig. 12.

**Fig. 4**. **Snapshots of ice concentrations in 3:2 resonance orbits**. Left column: coupled atmosphere-ocean experiments without sea-ice drift, and right column: fully coupled atmosphere-ocean-sea-ice experiments with sea-ice drift. Experimental design is the same as that in Fig. 1 except that the ratio of orbital period to rotation period is 3:2, rather than 1:1 (Supplementary Table 2). Sea ice dynamics reduce the open ocean area and lead TRAPPIST-1e to a snowball state and Kepler-442b to a nearly snowball state. The results of Kepler-1229b, Kapteyn b[*], TRAPPIST-1f, and Kepler-186f are similar to LHS 1140b, a snowball state, due to their relatively low stellar fluxes (< 700 W m$^{-2}$). For the details of how stellar flux, planetary radius, rotation period, surface albedo, and mixed layer depth influence the results, please see Supplementary Figs. 13 to 17.

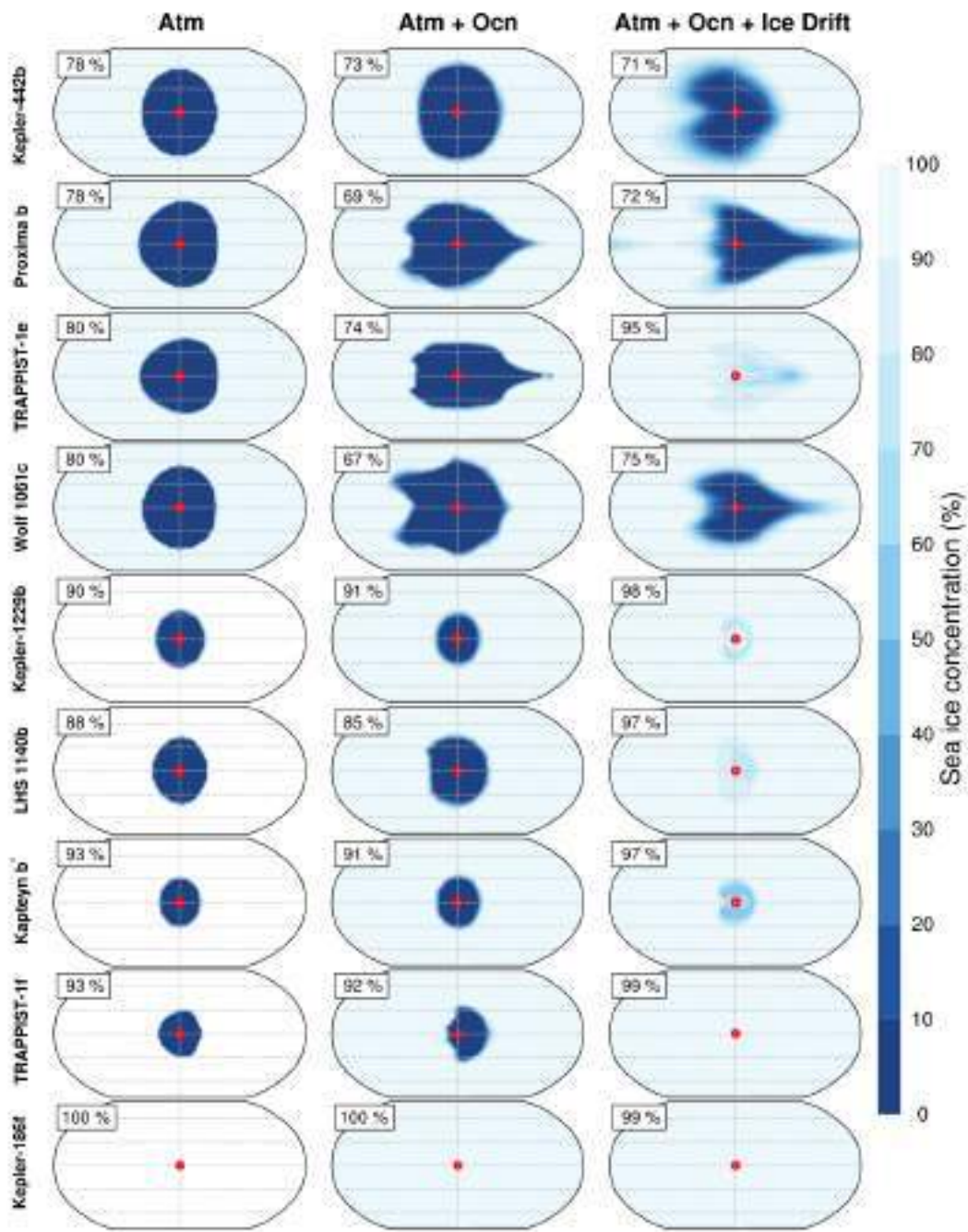

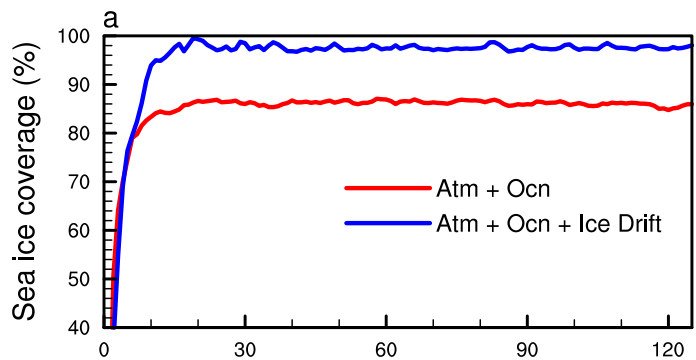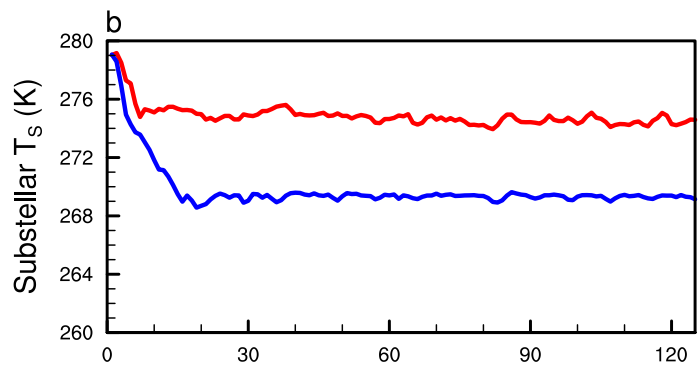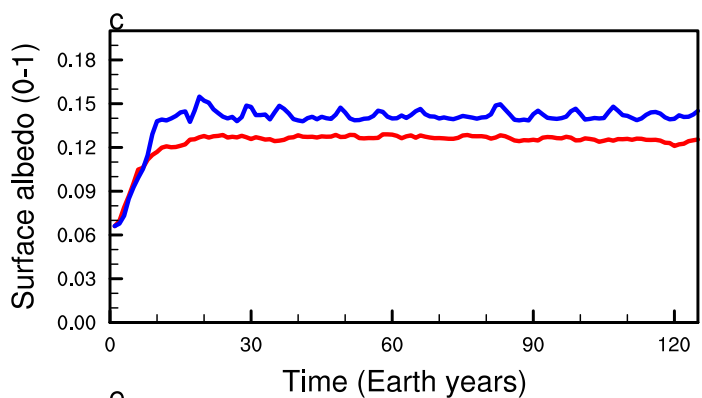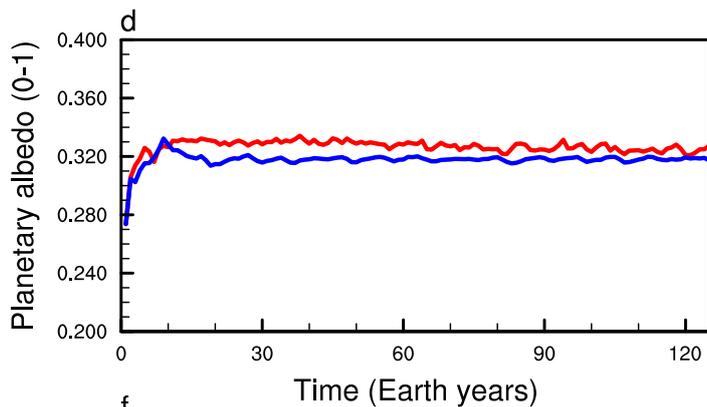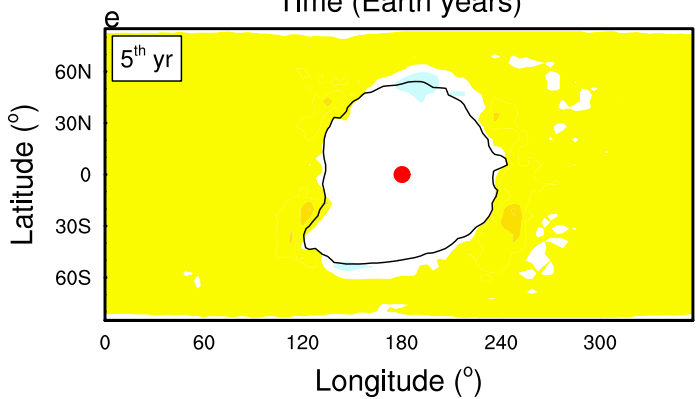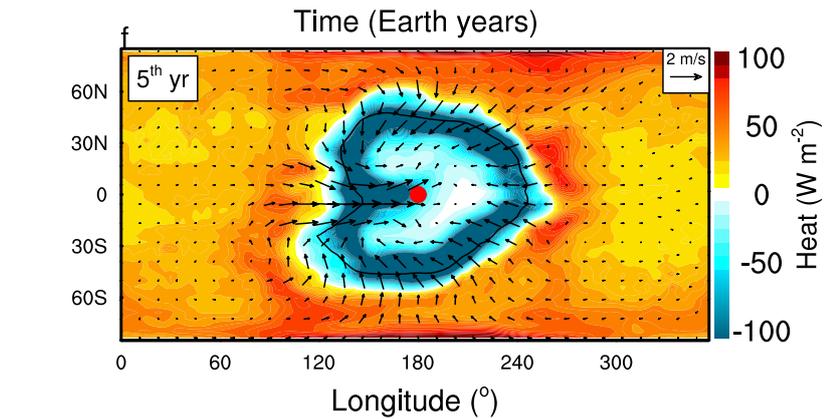

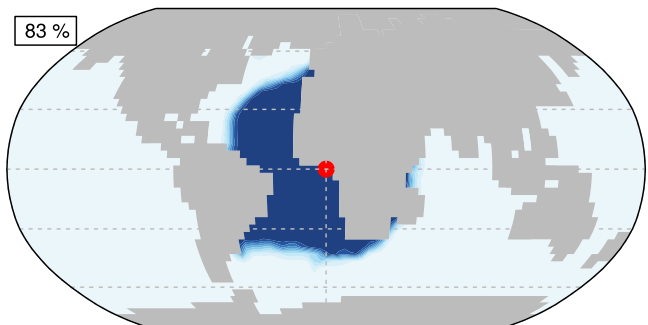
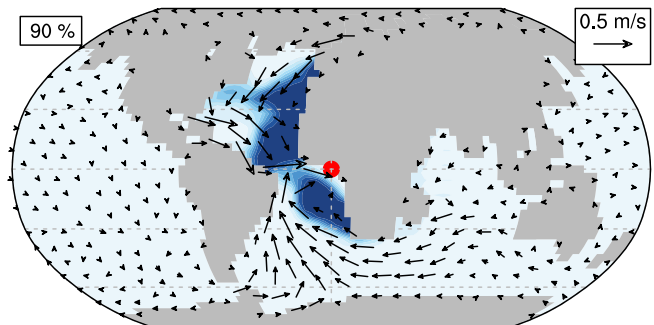
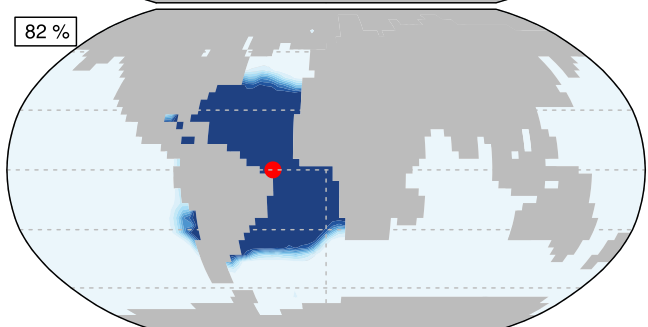
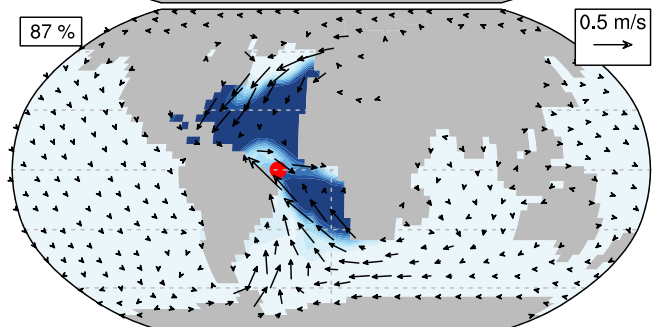
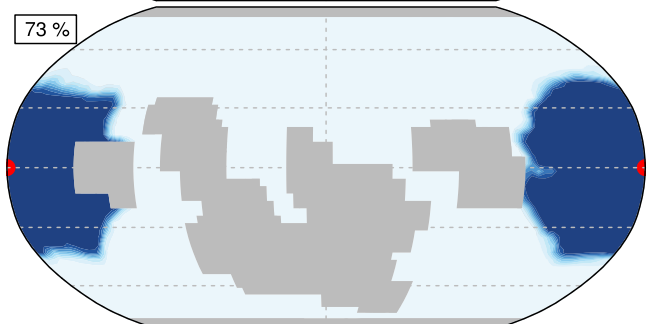
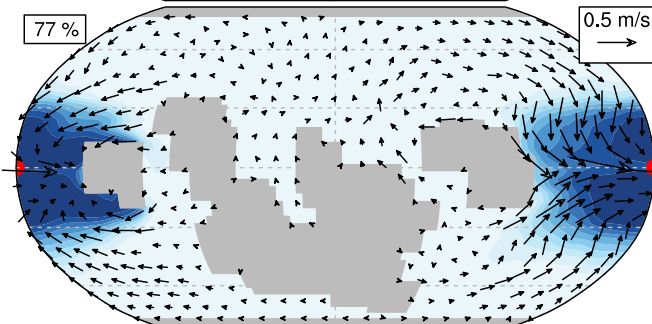
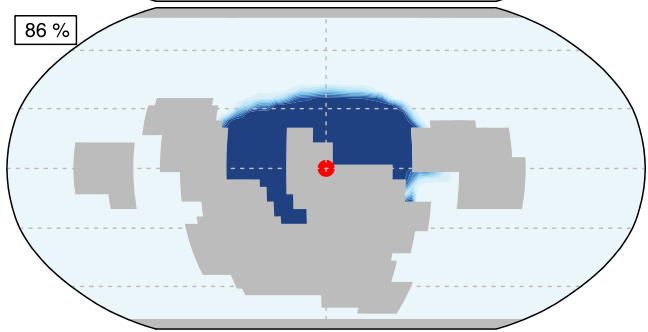
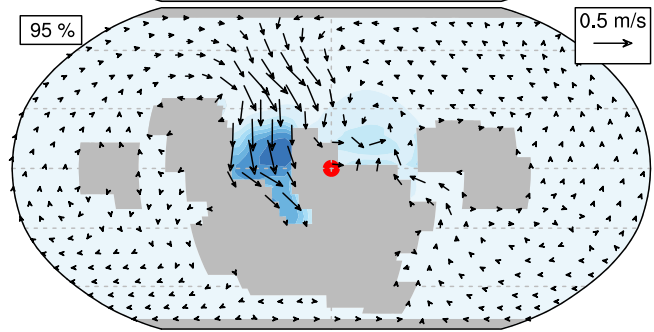
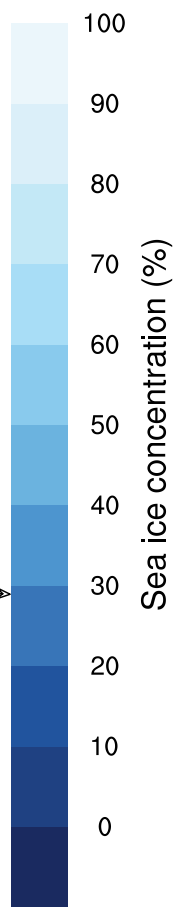

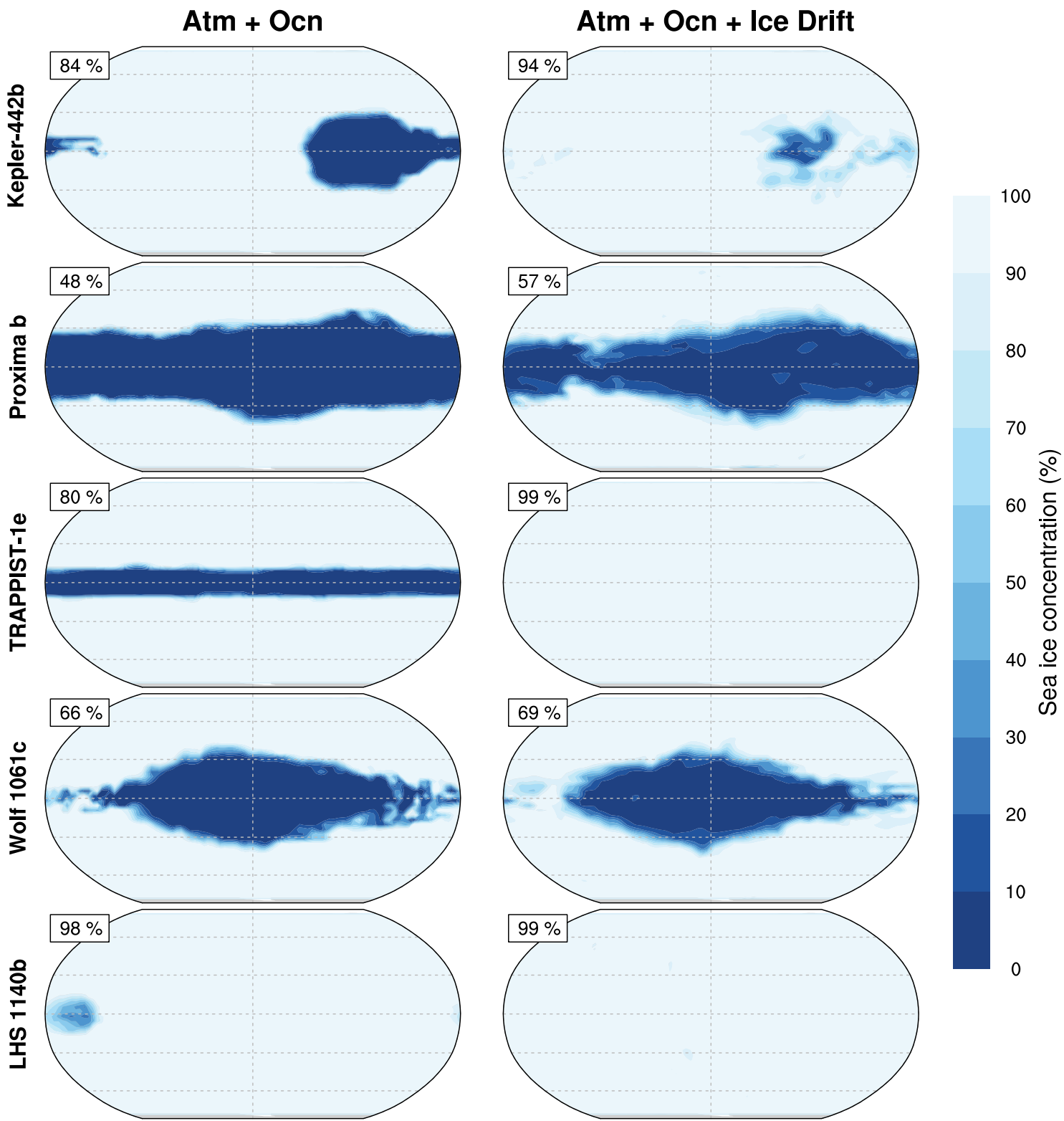

**Methods**
**Climate model and experimental design.** We employ a 3D global climate model, the Community Climate System Model version 3 (CCSM3). This model was developed by the National Center for Atmospheric Research of US in order to understand the climate of Earth in the present, past, and future[41,71]. We have modified the model to simulate the climates of tidally locked terrestrial planets in the habitable zone of low-mass stars[17,20,83,84]. The model includes four components—atmosphere, ocean, sea ice, and land, and one coupler is used to connect them. The atmosphere component is the 3D Community Atmospheric general circulation Model version 3, or called CAM3[40]. The ocean component is the 3D oceanic general circulation model—the Parallel Ocean Program (POP1)[74]. The land component is the Community Land Model (CLM3)[67]. The sea ice component is the Community Sea Ice Model (CSIM5), which simulates the sea ice in five thickness categories with four vertical layers in each category[37]. CSIM5 can run with thermodynamics only or with both thermodynamics and dynamics; this flexible makes us be able to explicitly examine the effect of sea ice dynamics on the climate.

In this work, three types of experiment are performed using CCSM3. The first is the atmosphere-only run coupled to a slab ocean. Both thermodynamics and dynamics of the atmosphere are considered. In the slab ocean, the thermo effects of the ocean and sea ice are considered but neither ocean dynamics or sea ice dynamics are included, i.e., both the ocean and sea ice are immobile. The second is the coupled atmosphere-ocean simulation with atmosphere thermodynamics and dynamics, ocean thermodynamics and dynamics, and sea ice thermodynamics, but without sea ice dynamics. The third is the fully coupled atmosphere-ocean-sea ice simulation with all of atmosphere, ocean, and sea ice thermodynamics and dynamics. We could estimate the effect of oceanic heat transport by comparing between the first and second groups of experiments and investigate the effect of sea-ice drift by comparing between the second and third groups of experiments.

A low resolution of CCSM3, labeled 'T31_gx3', is employed in this work. For Earth, this resolution is able to resolve large-scale atmospheric and oceanic circulations and marginally resolve mid-latitude mesoscale eddies in the atmosphere, but mesoscale eddies in the ocean are parameterized using the GM scheme[49]. For terrestrial planets in the habitable zone of M dwarfs, the Rossby deformation radius is larger than that on Earth due to the slow rotation rate, so that the model may be able to resolve mesoscale eddies in the ocean at least in the tropics[10,17]. The atmosphere and land components have the same horizontal resolution, 3.75° by 3.75° in meridional (South-North) and zonal (West-East) directions. The atmosphere has 26 vertical levels from the surface to ≈36 km, including the whole troposphere and the lower part of the stratosphere. The ocean and sea ice components have the same horizontal resolution, 0.9° near the equator and 3.6° near the poles in meridional direction and 3.6° (constantly) in the zonal direction. All the experiments are started from an ice-free state. Tests showed that the final equilibrium states do not depend on the initial state[83]. For tidally locked planets, there is nearly no climate bistability or hysteresis[28]. This is due to the fact that the stellar flux increases strongly as the substellar point is approached. Each experiment was run for tens, hundreds, or thousands of Earth years, mainly depending on the ocean depth. Note that in the snowball state the global-mean ice coverage is always a little smaller than 100% (see

Fig. 1), which may be due to that our integration period (always less than 1000 Earth years) is not long enough.

In order to test the effect of model resolution on the result, one additional experiment with a higher resolution was done, in which the atmosphere and land components have a horizontal resolution of 2.8° by 2.8°, the ocean and sea ice components have a resolution of 0.3° near the equator and 1.0° near the poles, and the ocean has 40 vertical levels, rather than 26 in the default resolution. Comparing this higher resolution experiment with the default experiment, the difference in global-mean sea ice coverage is less than 1%, and the spatial patterns are similar except that the higher resolution case has more fine structures near the ice edges (Supplementary Fig. 12).

**Comparisons between this study and previous studies.** Most of previous studies of tidally locked terrestrial planets employed 3D atmosphere-only global models coupled to a slab ocean[5-9,20,22,27,28,35,36,38,39,47,48,52,53,57-61,64-66,72,78,80-82,85]. Neither ocean dynamics nor sea ice dynamics have been considered in their work. Recent studies[10,17,79,83] found that fully coupled atmosphere-ocean-sea-ice processes are important for simulating the spatial pattern and thickness of sea ice on tidally locked planets. However, both ocean dynamics and sea ice dynamics are included in their simulations, so that it is unclear on the role of each process. Cullum et al.[42,43] examined the effects of planetary rotation rate and ocean salinity on the thermohaline circulation using an ocean-only model. Wind-driven ocean circulation and the interactions between ocean, atmosphere, and sea ice, however, were not considered in their simulations.

Here we clearly distinguish the effects of the two processes—ocean dynamics and sea ice dynamics, through turning on or off the module of sea ice dynamics in the model. Separating the effect of sea-ice drift from that of ocean dynamics is important. For example, Hu & Yang[17] showed that the stellar flux threshold for falling into a snowball state in the fully coupled atmosphere-ocean-sea-ice experiments is much higher than that in the atmosphere-only experiments (similar to that shown in Supplementary Fig. 1), however, the underlying mechanism is unclear because both ocean dynamics and sea ice dynamics can influence the sea ice coverage. In this paper, we clearly show that sea ice dynamics are the key mechanism whereas ocean dynamics have the opposite effect. Moreover, to our knowledge, no simulation with ocean dynamics or sea ice dynamics has been applied to TRAPPIST-1e and TRAPPIST-1f and no 3D global climate simulation has been applied to LHS 1140b, Wolf 1061c, Kepler-1229b, Kapteyn b, and Kepler-186f, so that our work presented here is the first exploration of the climates of these planets in the view of a coupled atmosphere-ocean system and a fully coupled atmosphere-ocean-sea-ice system. The work of Hu & Yang[17] focused on the unconfirmed planet—Gl 581g while Del Genio et al.[10] investigated the planet—Proxima b only. These two planets have relatively higher stellar fluxes (>850 W m$^{-2}$), so their climates are relatively warmer, which is not beneficial to uncover the effect of sea-ice drift as shown in this paper.

**Sea-ice drift and ice albedo.** The sea-ice drift equation is:
$$\rho \frac{\partial \vec{u}}{\partial t} = \vec{\tau_a} + \vec{\tau_o} - \rho f \vec{k} \times \vec{u} + \rho g \nabla H_o + \nabla \cdot \sigma_{ij}, \quad (1)$$

where ρ is the mean density of sea ice and snow over the ice, $\vec{u}$ is the sea ice velocity, $f$ is the Coriolis parameter, and $g$ is the planetary gravity. The right terms of the equation are the air-ice stress ($\vec{\tau_a}$), ocean-ice stress ($\vec{\tau_o}$), Coriolis force, pressure force associated with ocean surface slope ($H_o$), and internal ice stress ($\sigma_{ij}$), respectively. Non-linear advection terms are not considered in the model because they are negligible[37]. $\vec{\tau_a}$ and $\vec{\tau_o}$ are the main driven forces where the other terms are smaller on both Earth[56,76] and tidally locked planets[20]. For a given sea ice velocity, the Coriolis force in our simulations is smaller than that on Earth because of the slower rotation rates.

Oceanic heat transport and sea-ice flows act to significantly limit the sea ice thickness on tidally locked planets[10,17,20], so that the sea ice thickness is only several meters or tens of meters in the experiments with ocean and sea ice dynamics (Supplementary Fig. 9), and therefore sea glacier dynamics[30,68] are not required to consider here. After one planet enters a snowball state, ice thickness may reach one thousand meters or even thicker. This thick ice can flow under its own weight, but this process is beyond the scope of this paper. Here, we focus on the transition process from an ice-free state or a partially ice-covered state to a snowball state, during which the ice is thin.

The heat release (uptake) during sea ice growth (melting) is calculated following:
$$H = \rho_i L_i dh/dt, \qquad (2)$$
where $\rho_i$ is the ice density (917 kg m$^{-3}$), $L_i$ is the latent heat of fusion of ice (3.337 ×10$^5$ J kg$^{-1}$), and $dh/dt$ is the growth rate or the melting rate. Because of the very high value of $L_i$, a small growth or melting rate requires a large amount of energy. For example, a value of 2.8 cm day$^{-1}$ of $dh/dt$ is corresponding to an energy flux of 100 W m$^{-2}$.

For planets around M dwarfs, the ice albedo should be lower than that on Earth, due to the red shift of the stellar spectra compared to the Sun[21,72], except when a hydrohalite (NaCl·2H$_2$O) crust forms at low temperatures. Hydrohalite is much more reflective than bare ice or even snow in the near infrared[73]. Several sensitivity experiments with different ice albedos were performed (Supplementary Tables 1 & 2). For simplicity, snow albedo is assumed to be the same as the ice albedo. If a higher albedo were employed for snow, one planet may enter a snowball state at a higher stellar flux or CO$_2$ concentration threshold, but this does not influence our conclusion about sea ice dynamics.

**Planetary properties.** Supplementary Table 1 lists the observed parameters of the nine planets (ref. https://en.wikipedia.org/wiki/List_of_potentially_habitable_exoplanets), as well as Earth's values for reference. All of them are likely to have a rocky composition ($0.5R_\oplus \leq R_p \leq 1.6\ R_\oplus$ and $0.1\ M_\oplus \leq M_p \leq 6\ M_\oplus$, where $R_p$ is the planetary radius and $M_p$ is the planetary mass,) and are in the outer regions of the liquid water habitable zone around low-mass stars ($S_p \leq 0.7 S_\oplus$, where $S_p$ is the stellar flux at the substellar point and $S_\oplus$ is that on Earth, 1365 W m$^{-2}$). We use the incident stellar spectra from the BT-Settl stellar model[33]. The planets are close to the host stars, so that they may be in synchronous rotation orbit (i.e., rotation period = orbital period, with permanent day and night), or in spin—orbit resonance if orbital eccentricity is non-zero[34] or if the effect of thermal tides is significant[62]. Here, we examine two types of orbit, a synchronous rotation and a 3:2 spin—orbit resonance. There are seven planets in the TRPPIST-1 system; here, we focus on TRAPPIST-1e and TRAPPIST-1f, which are potentially habitable. The other five planets in the TRPPIST-1 system are not in the habitable zone[7,9].

Note that there are significant uncertainties in the observed parameters. For example, Gillon *et al.*[3] reported that the stellar flux for TRAPPIST-1e is 0.662 $S_⊕$ (used in E. T. Wolf's simulations[7]), however, new observations[44] with the *Spitzer Space Telescope* found that the stellar flux may be 0.604 $S_⊕$ (used in this study). Moreover, recent *Gaia* mission data shown that after the updates of star distance the stellar flux may be 5.5% higher on TRAPPIST-1e and 44% higher on LHS 1140b than previous estimates[54]. In order to investigate how large the uncertainty in the stellar flux could influence the result, we perform an additional experiment for TRAPPIST-1e with a stellar flux of 904 W m$^{-2}$, comparing to the default value of 821 W m$^{-2}$. We find that the global-mean sea ice coverage decreases from a snowball state to 83%, in the fully coupled experiments. There are also some uncertainties in planetary mass and radius (and thereby the estimated gravities)[44,50]. These uncertainties, however, do not influence our conclusion about sea ice dynamics under given planetary parameters.

In order to test the effects of stellar flux, planetary rotation period, size, and gravity on the sea ice coverage, we have also performed several series of sensitivity experiments. The tested parameters (Supplementary Table 2) almost cover the whole range of possible tidally locked terrestrial planets in outer regions of the habitable zone of M dwarfs. Geothermal heat flux is zero in the experiments.

**Atmospheric mass and greenhouse gas concentration.** Atmospheres of the planets are unknown since present telescopes are still unable to measure the atmospheric compositions of Earth-size planets in the habitable zone of M dwarfs. Photolysis models[45,69] have been employed to examine this problem and found that many atmospheric scenarios are possible, including airless with no atmosphere, $CO_2$-dominated atmosphere (similar to Mars and Venus), $N_2$-dominated atmosphere (similar to Earth and Titan), $O_2$-dominated atmosphere after the escape of H during the pre-main sequence of the host stars, etc. In this work, we assume the atmosphere is similar to modern Earth or early Earth, $N_2$-dominated with $CO_2$, $CH_4$, and $H_2O$, same as previous studies[7,8,10]. Different background air masses have been tested, $3.1 × 10^3, 1.0 × 10^4$ (the default one), $2.0 × 10^4, 3.1 × 10^4$, and $1.0 × 10^5$ kg m$^{-2}$ in column mass, and the corresponding surface pressures are respectively 1.01, 3.24, 6.28, 10.1, and 32.4 bar under for example LHS 1140b's gravity of 31.8 m s$^{-2}$. When compare the climate differences among the planets, we employ the same column air mass rather than surface pressure because the surface pressure depends on gravity.

By default, atmospheric $CO_2$ mixing ratio is 300 ppmv (parts per million by volume), $CH_4$ 0.8 ppmv, and $N_2O$ 0.27 ppmv, similar to modern Earth. High concentrations of greenhouse gases, similar to early Earth, have also been considered, such as 1000 ppmv of $CH_4$ and 1200, 4800, 19,200, 76,800, and $10^5$ ppmv of $CO_2$. Of course, an even higher concentration of $CO_2$ is possible, which may be able to melt all the sea ice[17] and then sea ice dynamics have no any effect. For Proxima b, lower concentrations of $CO_2$, 3 and 30 ppmv, were examined; an experiment with no any greenhouse gas (zero $CO_2$, $CH_4$, and N₂O) was also performed, in order to know whether Proxima b would enter a snowball state when there is no any greenhouse gas (Supplementary Fig. 8). When investigate the effect of varying air mass, the greenhouse gas mass (rather than mixing ratio) is set to the same, such as the $CO_2$ mixing ratio is 300 and 30 ppmv for the column air masses of $1.0 \times 10^4$ and $1.0 \times 10^5$ kg m$^{-2}$, respectively, so that in these two sets the column $CO_2$ mass is the same, ≈4.5 kg m$^{-2}$. Note that the greenhouse effect of $CO_2$, as well as other greenhouse gases, is determined by absolute molecular number.

As addressed above, we have considered several different atmospheric compositions, but it is still very limited mainly due to the limitation of the radiative transfer module of the model. However, our work is in order to examine the effect of sea ice dynamics on the climate and we believe that our experiments are enough to reach this target.

**Land-sea distribution and ocean depth.** Several different land-sea distributions have been used in this work, an aqua-planet, modern Earth, 630-Ma Earth, an idealized super-continent in the tropics, and two idealized super-continents. The aqua-planet has no continent except that two small islands are added to the south and north poles because the poles of the ocean grid have to reside on continents[71]. Due to computational limitations, the ocean depth is set to be 1000 m by default for the aqua-planet experiments, so that only wind-driven ocean circulation is considered but not for the deep ocean circulation that requires much longer integration time. Two additional experiments with an ocean depth of 4000 m (close to the mean value of Earth) are performed. For the configurations with continents, the ocean depth is set to be 4000 m by default, except that in the modern Earth configuration realistic ocean depths are used. The substellar point locates over the ocean or the land. Both seawater salinity and sea ice salinity are set to 4 g kg$^{-1}$, in order to minimize the effect of haline circulation, which also requires long integration time. The freezing point is a constant, $-1.8°C$.

In atmosphere-only experiments, a slab ocean of 50 m (by default) was used. For a 1:1 orbit, the mixed layer depth does not affect the final equilibrium state[22]. For a 3:2 resonance orbit, four different depths of the slab ocean, 1, 5, 10, and 350 m, were tested additionally (Supplementary Fig. 17). One could find the mixed layer depth can influence the results significantly. The value of 350 m is close to the time mean mixed layer depth of the deep tropics internally calculated in the coupled experiments.

**Atmospheric reflection.** Supplementary Figure 10 shows that the surface climate nearly does not depend on sea ice albedo when surface albedo is less than atmospheric reflection (R). R is contributed from the combined effect of atmospheric scattering and cloud albedo and is equal to:

$$R = \frac{SF_{TOA}^{\uparrow} - \alpha_s (F_{SURF}^{\downarrow})^2}{S^2 - \alpha_s^2 (F_{SURF}^{\downarrow})^2}, \qquad (3)$$

where $\alpha_s$ is the surface albedo, $S$ is the incoming stellar flux at the top of the atmosphere (TOA), $F_{TOA}^{\uparrow}$ is the upwelling stellar flux at the TOA, and $F_{SURF}^{\downarrow}$ is the downwelling stellar flux at the surface. We used model outputs of these variables ($\alpha_s$, $S$, $F_{TOA}^{\uparrow}$, and $F_{SURF}^{\downarrow}$) to calculate $R$. This equation has considered the effect of multiple reflections between the surface and the atmosphere (ref. Fig. 1 in Donohoe & Battisti[46]). Both the atmosphere and surface contribute to planetary albedo ($\alpha_p$). The atmospheric contribution is $R$ while the surface contribution is equal to $\alpha_p - R$. The value of $\alpha_p - R$ is always smaller than $\alpha_s$, because atmospheric processes attenuate the surface contribution[46]. For tidally locked planets, this attenuating effect is more significant than that on Earth, because the dayside is covered by optically thick clouds[22,59,79].

**Rossby deformation radius.** In understanding the atmospheric circulation on tidally locked planets, a characteristic factor is the ratio of the equatorial Rossby deformation radius ($L_R$)[18,85] to planetary radius (a):

$$\frac{L_R}{a} \equiv \sqrt{\frac{NH}{\beta a^2}} = \sqrt{\frac{NH}{2\omega a}} = \left(\frac{R^{*2}T}{M^2 C_p} - \frac{R^*T}{M}\frac{\partial lnT}{\partial lnp}\right)^{\frac{1}{4}} / (2\omega a)^{\frac{1}{2}}, \quad (4)$$

where $N$ is the Brunt-Vaisala frequency in the troposphere, $H$ is the scale height, $\beta$ is the derivative of the Coriolis parameter with northward distance at the equator (equal to $2\omega/a$), ω is the planetary rotation rate, $R^*$ is the universal gas constant, $M$ is the molecular molar mass, $T$ is the air temperature, $p$ is the air pressure, and $C_p$ is the specific heat at constant pressure. Note that the value of $L_R/a$ does not directly depend on gravity. In Supplementary Fig. 4f, the slight decrease of $L_R/a$ with increasing the gravity is mainly due to the small changes of stratification ($\partial lnT/\partial lnp$).

**Comparing the effects of varying rotation period and varying radius.** From Supplementary Figure 4, one could find that ocean dynamics act to decrease the sea ice coverage whereas sea ice dynamics have the opposite effect, and that the global-mean ice coverage is insensitive to vary the rotation period (as well as gravity) in all the three types of run and to vary the planetary radius in the atmosphere-only experiments, but it is sensitive to vary the radius in the coupled and fully coupled experiments.

For large-scale atmospheric circulation, the effect of increasing the radius should be similar to that of decreasing the rotation period, due to the fact that these two parameters influence the value of $L_R/a$ in the same way (Eq. (4)). Indeed, both the strength and spatial pattern of zonal-mean zonal winds in the experiments of doubling planetary radius (from 1 to 2 $R_e$) are similar to those in the experiments of halving rotation period (from 10 to 5 days) in the atmosphere-only runs (Supplementary Fig. 6). However, this equivalence breaks up in the coupled atmosphere-ocean runs. In the coupled experiments, the case of $2R_e$ exhibits one westerly jet in the tropics and one easterly jet in the sub-polar region of each hemisphere but the experiment of 5 days has one middle-latitude westerly jet on each hemisphere; the easterly jets are connecting to westward and polarward (due to the Coriolis effect) surface winds, which transport heat toward to the polar sea ice and melt the ice there (Supplementary Fig. 5). These different responses in atmospheric circulation are due to that the effect of varying the radius on the oceanic circulation is different from that of varying the rotation period; atmosphere-ocean interactions further change the atmospheric circulation.

Varying rotation period influences the Ekman layer depth but varying radius does not. In theory, the Ekman layer depth[77] is equal to $\sqrt{2A_z/f}$, where $A_z$ is vertical mixing coefficient and $f$ is the Coriolis parameter. In steady state, the zonal momentum equation in the primitive equations of oceanic motion[75] is:

$$\frac{u}{fa\cos\theta}\frac{\partial u}{\partial \lambda} + \frac{v}{fa}\frac{\partial u}{\partial \theta} + \frac{w}{f}\frac{\partial u}{\partial z} - v = -\frac{1}{fa\rho\cos\theta}\frac{\partial p}{\partial \lambda} + \frac{1}{f}\frac{\partial}{\partial z}\left(A_z \frac{\partial u}{\partial z}\right), \quad (5)$$

where $\lambda$ is longitude, $\theta$ is latitude, $a$ is the planetary radius, and other variables are standard. For concision, metric terms and horizontal viscosity terms are not written in this equation. From this equation, it is clear to sea that for the first and second terms on the left side and the first term on the right side, increasing $f$ has an equivalent effect to that of increasing $a$, however, changing $f$ can influence the third term on the left side and the mixing term on the right side but changing $a$ cannot. Note that varying the radius influences the horizontal scale only, and it does not affect the vertical scale. If all other things being equal, when $f$ is increased, the Ekman layer becomes shallower especially in middle latitudes where $f$ is relatively greater (Supplementary Fig. 5g), so that less heat could be stored in the upper ocean and then be transported to other regions; increasing $a$ has no this effect. Moreover, in sea-ice drift, the effect of varying $a$ is also different from that of varying $f$. This is due to the fact that the ratio of the distance of sea-ice drift to planetary radius, $u_i t/a$, decreases with increasing $a$ for a given ice velocity ($u_i$) under a given time ($t$). Varying $f$ has no this effect. Future work is required to more clearly examine the effect of varying $a$ on the climate using ocean-only and ice-only models.

**The effect of an active carbon cycle.** $CO_2$ concentration is specified in our simulations, so that the effect of the carbon-silicate cycle is not considered in this study. Previous studies have shown that for planets lacking land vascular plants and located in the outer regions of the habitable zone, their climates may oscillate between long periods of snowball state and shorter periods of warm state, on geological timescales[27,31,55]. This is due to the direct dependence of weathering rate on $CO_2$ partial pressure and precipitation rate. Moreover, for synchronously rotating planets around low-mass stars and for planets with long solar days, recent studies[28,32] found that the period of a snowball state would be very short or the planets could equilibrate with a small ice-free ocean centered on the substellar point. This is due to the special spatial pattern of the stellar radiation—the strong increase in instellation as the substellar point is approached. For synchronously rotating planets, the transition between partial and complete ice coverage is smooth rather than abrupt, i.e., no runaway glaciation bifurcation. Our results presented here suggest that the period of a snowball state would be longer than that suggested in Checlair *et al.*[28], or it requires a higher $CO_2$ concentration or a larger $CO_2$ outgassing rate to prevent the planets entering a snowball state. For planets with an inactive carbon cycle, whether they would fall into a snowball state depends on their initial $CO_2$ concentrations. For either active or inactive carbon cycle, sea ice dynamics would increase the $CO_2$ concentration threshold for the onset of a snowball state.

**Data availability** The data that support the plots within this paper and other findings of this study are available from the corresponding author upon reasonable request.

**Code availability** The source codes of the model CCSM3 can be downloaded from www.cesm.ucar.edu/models/ccsm3.0 and changes of the model are available from the corresponding author.

# Supplementary Information
for the paper "Transition from Eyeball to Snowball Driven by Sea-ice Drift on Tidally Locked Terrestrial Planets" by Jun Yang *et al.*, 2019

**Contents:**

1. Supplementary Tables 1 & 2
2. Supplementary Figures 1 to 19
3. Supplementary Video 1

**Supplementary Table 1. Planetary properties**, including star temperature, stellar flux at the substellar point, orbital period, orbital period : rotation period, planetary radius, gravity, and ice albedo in the visible band and the near-infrared band**.** The planets are listed in order of decreasing stellar flux.

| Planets | Stellar flux (W m$^{-2}$) | Star (K) | Orbital period (Earth days) | Orbit : Rotation | Radius (10$^6$ m) | Gravity (m s$^{-2}$) | Ice albedo (0-1) |
|---|---|---|---|---|---|---|---|
| Earth | 1365 | 5789 | 365 | 365:1 | 6.4 | 9.8 | 0.80, 0.50 |
| Kepler-442b | 956 | 4402 | 112 | 1:1 or 3:2 | 9.0 | 12.5 | 0.75, 0.30 |
| Proxima b | 887 | 3042 | 11.2 | 1:1 or 3:2 | 8.1 | 12.0 | 0.65, 0.15 |
| TRAPPIST-1e | 821 | 2516 | 6.1 | 1:1 or 3:2 | 5.8 | 9.1 | 0.60, 0.10 |
| Wolf 1061c | 819 | 3342 | 17.9 | 1:1 or 3:2 | 10.2 | 16.7 | 0.70, 0.20 |
| Kepler-1229b | 669 | 3784 | 86.8 | 1:1 or 3:2 | 8.9 | 13.7 | 0.75, 0.25 |
| LHS 1140b | 625 | 3100 | 24.7 | 1:1 or 3:2 | 9.1 | 31.8 | 0.65, 0.15 |
| Kapteyn b$^*$ | 587 | 3550 | 48.6 | 1:1 or 3:2 | 10.2 | 18.4 | 0.70, 0.20 |
| TRAPPIST-1f | 519 | 2516 | 9.2 | 1:1 or 3:2 | 6.7 | 8.3 | 0.60, 0.10 |
| Kepler-186f | 400 | 3755 | 130 | 1:1 or 3:2 | 7.5 | 10.8 | 0.75, 0.25 |

**Supplementary Table 2. Summary of the climate experiments**

| Group | Runs | Experimental Design |
|---|---|---|
| Control | 27 | Control experiments of the nine planets. Three types of experiment for each planet: atmosphere-only, coupled atmosphere-ocean, and fully coupled atmosphere-ocean-sea-ice. The surface is a water world with an ocean depth of 1000 m. In the atmosphere-only experiments, a slab, immobile ocean of 50 m is used. Rotation period is equal to orbital period. Column air mass is set to be the same, $1.0 \times 10^4$ kg m$^{-2}$ (N$_2$) with 300 ppmv CO$_2$ (corresponding to a column CO$_2$ mass of ≈4.5 kg m$^{-2}$ under this air mass), 0.8 ppmv CH$_4$, and 0.27 ppmv N$_2$O. Figures 1 & 2, and Supplementary Figures 2, 9, & 19, and Video 1. |
| Stellar flux | 24 | Examining the effects of oceanic heat transport and sea ice drift on the stellar flux threshold for the onset of a snowball glaciation. The stellar fluxes tested are 400, 500, 550, 600, 700, 800, 850, 900, and 1000 W m$^{-2}$. All three types of experiment were performed. A star temperature of 3000 K, 1.0 Earth's radius, 1.0 Earth's gravity, 10-day rotation period (= orbital period). Column air mass is $1.0 \times 10^4$ kg m$^{-2}$ (N$_2$), and the column CO$_2$ mass is 4.5 kg m$^{-2}$. The sea ice albedo is 0.15 in the near infrared and 0.65 in the visible. Supplementary Figure 3. |
| Air mass | 10 | Examining the effect of background air (N$_2$) mass. Column air masses are $3.1 \times 10^3$, $1.0 \times 10^4$, $2.0 \times 10^4$, $3.1 \times 10^4$, and $1.0 \times 10^5$ kg m$^{-2}$. Column CO$_2$ mass is the same, 4.5 kg m$^{-2}$, and the corresponding volume mixing ratios are 967, 300, 150, 97, and 30 ppmv, respectively. LHS 1104b's parameters are used. Supplementary Figures 3d & 18. |
| Ice albedo | 10 | Examining the effect of sea ice albedo. The ice albedo is 0.65 in the visible band and 0.15 in the near-infrared band, labeled (0.65, 0.15), and five other designs: (0.70, 0.20), (0.80, 0.30), (0.80, 0.40), (0.80, 0.50), and (0.80, 0.60). Planetary parameters of Proxima b were employed. Two types of experiment were run, atmosphere-only and fully coupled atmosphere-ocean-sea-ice. Supplementary Figure 10. |
| Land-sea distribution | 18 | Examining the effect of land-sea distribution. Four configurations are employed. The ocean depth is 4000 m (uniform, close to the mean value of modern Earth) or the realistic values of modern Earth. Parameters of TRAPPIST-1e are employed. Column air mass is $1.0 \times 10^4$ kg m$^{-2}$ (N$_2$). Two types of experiment were run, coupled atmosphere-ocean and fully coupled atmosphere-ocean-sea-ice. Figure 3 and Supplementary Figure 11. |
| Greenhouse gas | 9 | Examining the effect of greenhouse gas. For Proxima b, three concentrations are performed, 30 ppmv of CO$_2$, 3 ppmv of CO$_2$, and zero CO$_2$, CH$_4$, and N$_2$O; for TRAPPIST-1e, 1200, 4800, and 19,200 ppmv of CO$_2$; for LHS 1140b, 19,200 ppmv of CO$_2$, 76,800 ppmv of CO$_2$, and 76,800 ppmv of CO$_2$ and 1000 ppmv of CH$_4$. By default, CH$_4$ is 0.8 ppmv and N$_2$O 0.27 ppmv. Supplementary Figure 8. |
| Rotation period | 21 | Examining the effect of rotation period. Same as the 'stellar flux' experiments, except that the rotation periods tested are 2.5, 5, 10, 15, 20, 25, and 30 Earth days. The stellar flux is 800 W m$^{-2}$. All three types of experiment were performed. Supplementary Figures 4, 5, & 6. |
| Radius | 18 | Examining the effect of planetary radius. Same as the 'stellar flux' experiments, except that the radii tested are 0.75, 1.0, 1.25, 1.5, 1.75, and 2.0 Earth's value. The stellar flux is 800 W m$^{-2}$. All three types of experiment were performed. Supplementary Figures 4, 5, & 6. |

**Supplementary Table 2 continued**

| Group | Runs | Experimental Design |
|---|---|---|
| Gravity | 15 | Examining the effect of planetary gravity. Same as the 'stellar flux' experiments, except that the gravities tested are 4.9, 9.8, 19.6, 29.4, and 34.3 m s$^{-2}$. The stellar flux is 800 W m$^{-2}$. All three types of experiment were performed. Supplementary Figure 4. |
| Internal ice stress | 2 | Examining the effect of internal ice compressive stress (P*). Same as the fully coupled control run of TRAPPIST-1e, except that two additional values of P* are tested, 5 and 100 kPa. Supplementary Figure 8d. |
| Stellar flux uncertainty | 3 | Examining the effect of stellar flux uncertainty in observations. A higher stellar flux, 904 W m$^{-2}$, is used for TRAPPIST-1e. Other parameters are unchanged. All three types of experiments were run. |
| Wolf 1061c | 4 | Examining the effects of varying planetary radius and rotation period on the climate of Wolf 1061c under the 1:1 orbit. Planetary parameters of Wolf 1061c are employed, except that the radius or rotation period of TRAPPIST-1e replaces the default one. Two types of experiment were run, coupled atmosphere-ocean and fully coupled atmosphere-ocean-sea-ice. Supplementary Figure 7. |
| Resolution | 2 | Examining the effect of model resolution on the result. We test a higher resolution, in which the resolution of atmosphere and land components is about 50% higher than the default one and the resolution of ocean and sea ice is about 200% higher than the default one. Planetary parameters of TRAPPIST-1e are employed. Supplementary Figure 12. |
| 3:2 resonance | 18 | Same as the 'control' experiments except for a 3:2 spin-orbital resonance state. Rotation period is 2/3 of orbital period. Both obliquity and eccentricity are zero. Figure 4. |
| Stellar flux_3:2 | 10 | Examining the effect of sea-ice drift for different stellar fluxes under a 3:2 resonance state. Parameters of Proxima b and Wolf 1061c are employed, except that the stellar flux is reduced. Two types of experiment were run, coupled atmosphere-ocean and fully coupled atmosphere-ocean-sea-ice. Supplementary Figure 13. |
| Rotation period_3:2 | 8 | Examining the effect of varying rotation period on the surface climate under 3:2 resonance states. Parameters of Proxima b are employed, except that the rotation period and orbital period are changed. Two types of experiment were run, coupled atmosphere-ocean and fully coupled atmosphere-ocean-sea-ice. Supplementary Figure 14. |
| Kepler-442b_3:2 | 4 | Examining the effects of varying surface albedo and rotation period on the climate of Kepler-442b under 3:2 resonance states. Parameters of Kepler-442b are employed, except that the surface albedos are reduced or the rotation period and orbital period are reduced. Two types of experiment: coupled atmosphere-ocean and fully coupled atmosphere-ocean-sea-ice. Supplementary Figure 15. |
| Radius_3:2 | 4 | Examining the effect of varying planetary radius on the climates of Proxima b and Wolf 1061c under 3:2 resonance orbits. Parameters of these two planets are employed, except that the radii are reduced to that of TRAPPIST-1e. Two types of experiment: coupled atmosphere-ocean and fully coupled atmosphere-ocean-sea-ice. Supplementary Figure 16. |
| Mixed layer depth | 10 | Examining the effect of ocean mixed layer depth on the result. Same as the atmosphere-only experiments of the '3:2 resonance', except that the mixed layer depths tested are 1, 5, 10, 50, and 350 m. Only Proxima b and TRAPPIST-1e were performed. Supplementary Figure 17. |

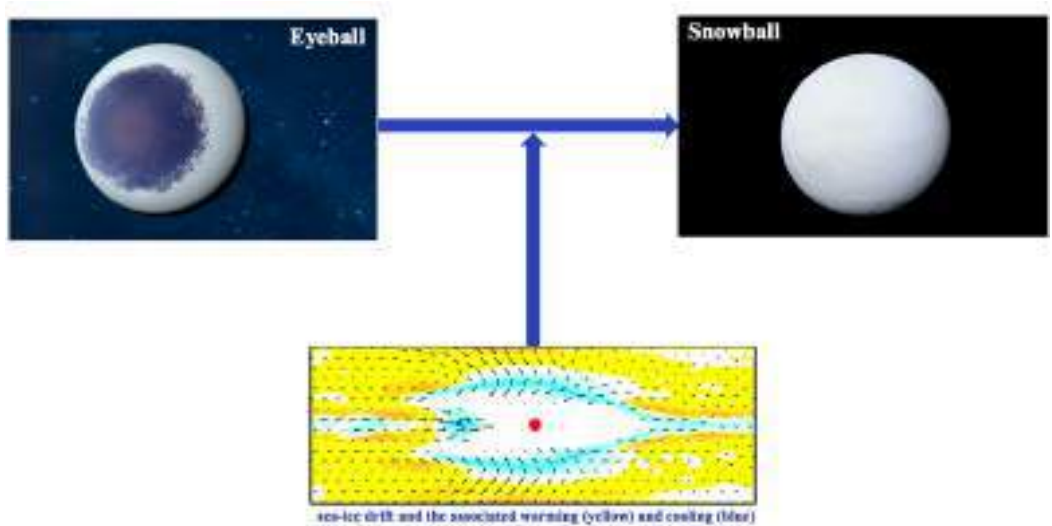

**Supplementary Figure 1.** Schematic illustration of the transition from an eyeball-like climate to a snowball climate driven by sea-ice drift on tidally locked terrestrial planets. Sea ice grows at the nightside and high latitudes of the dayside, drifts toward the relatively warmer substellar region, and a part of the ice melts. These processes warm the ice growth regions (yellow) but dramatically cool the melting regions (light blue), promoting the onset of a snowball state.

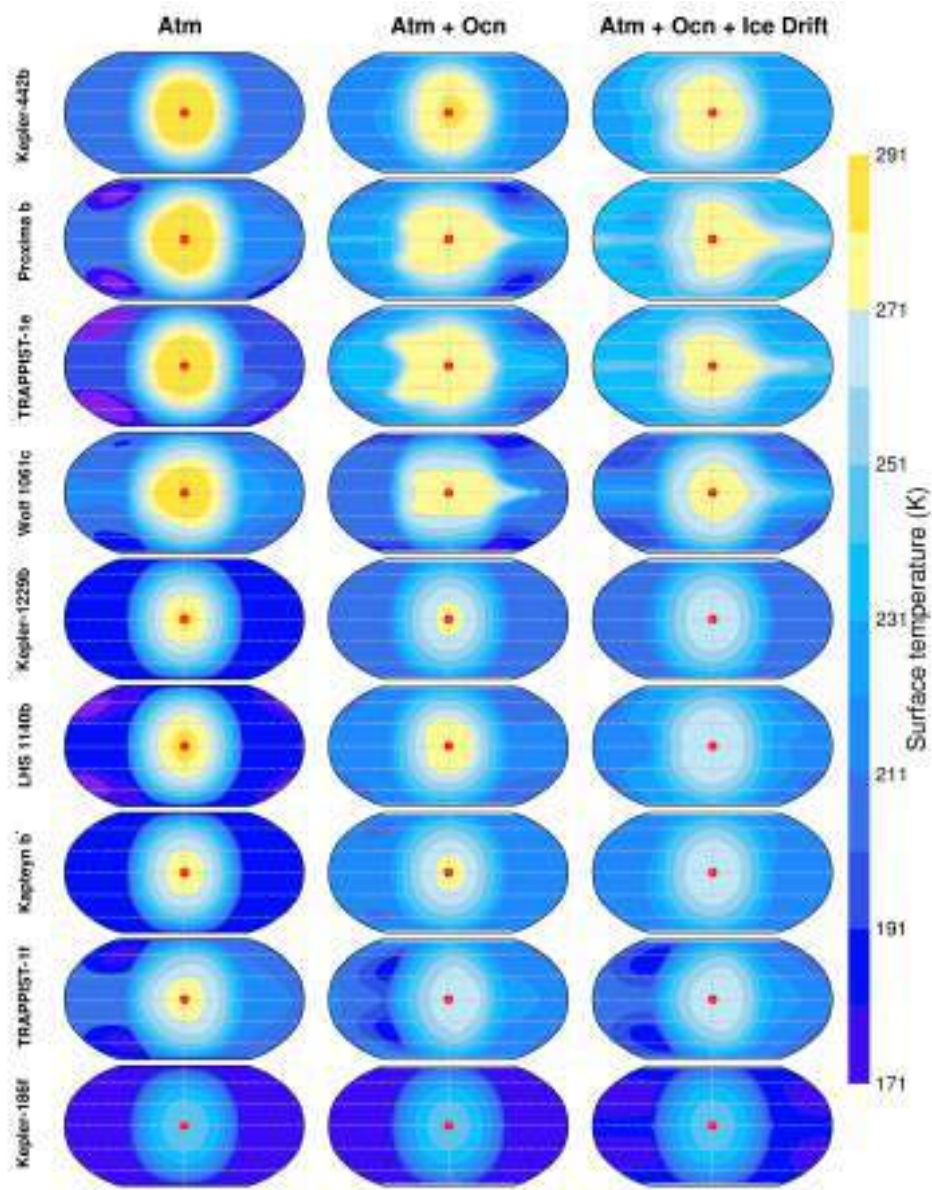

**Supplementary Figure 2**. Surface air temperatures in synchronous rotation orbits. This figure is the same as Fig. 1 shown in the main text, but for surface air temperatures. The freezing point is 271.35 K, or -1.8°C.

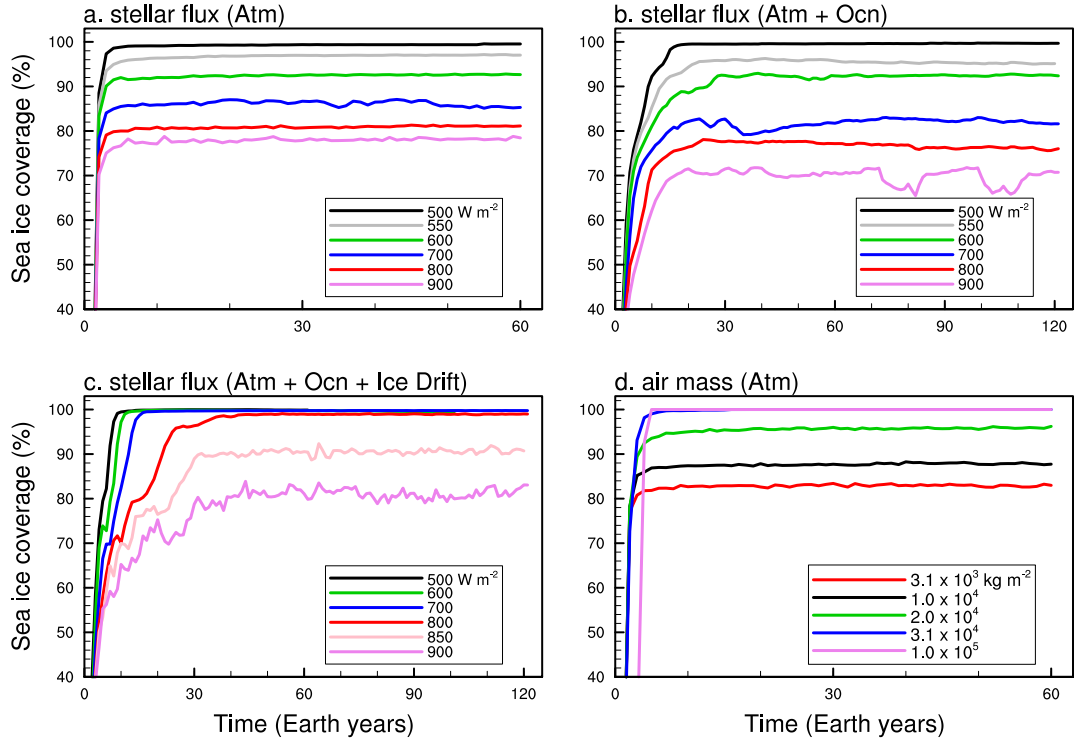

**Supplementary Figure 3**. Effects of ocean dynamics, sea ice dynamics, and air mass on the onset of a snowball glaciation. Time-series of global-mean sea ice coverage for different stellar fluxes from 500 to 900 W m$^{-2}$ in increments of 50 or 100 W m$^{-2}$ (**a-c**) and different column air (N$_2$) masses of $3.1 \times 10^3$, $1.0 \times 10^4$, $2.0 \times 10^4$, $3.1 \times 10^4$, and $1.0 \times 10^5$ kg m$^{-2}$ (**d**). In **a-c**, planetary parameters are the same, a 10-day rotation period (= orbital period), 1.0 Earth's radius, 1.0 Earth's gravity, and a column air mass of $1.0 \times 10^4$ kg m$^{-2}$. The planet enters a snowball state in a stellar flux of $\leq$ 500 W m$^{-2}$ in both the atmosphere-only experiments (**a**) and the coupled atmosphere-ocean experiments (**b**) but $\leq$ 800 W m$^{-2}$ in the fully coupled atmosphere-ocean-sea-ice experiments (**c**). In **d**, they are atmosphere-only experiments and the parameters of LHS 1140b were employed. In all of these experiments, the column CO$_2$ mass is set to be the same, $\approx$4.5 kg m$^{-2}$, and the initial condition is ice-free. Ocean dynamics expand the open ocean area but sea ice dynamics reduce the open ocean area, and a high background air mass also promotes the onset of a snowball state.

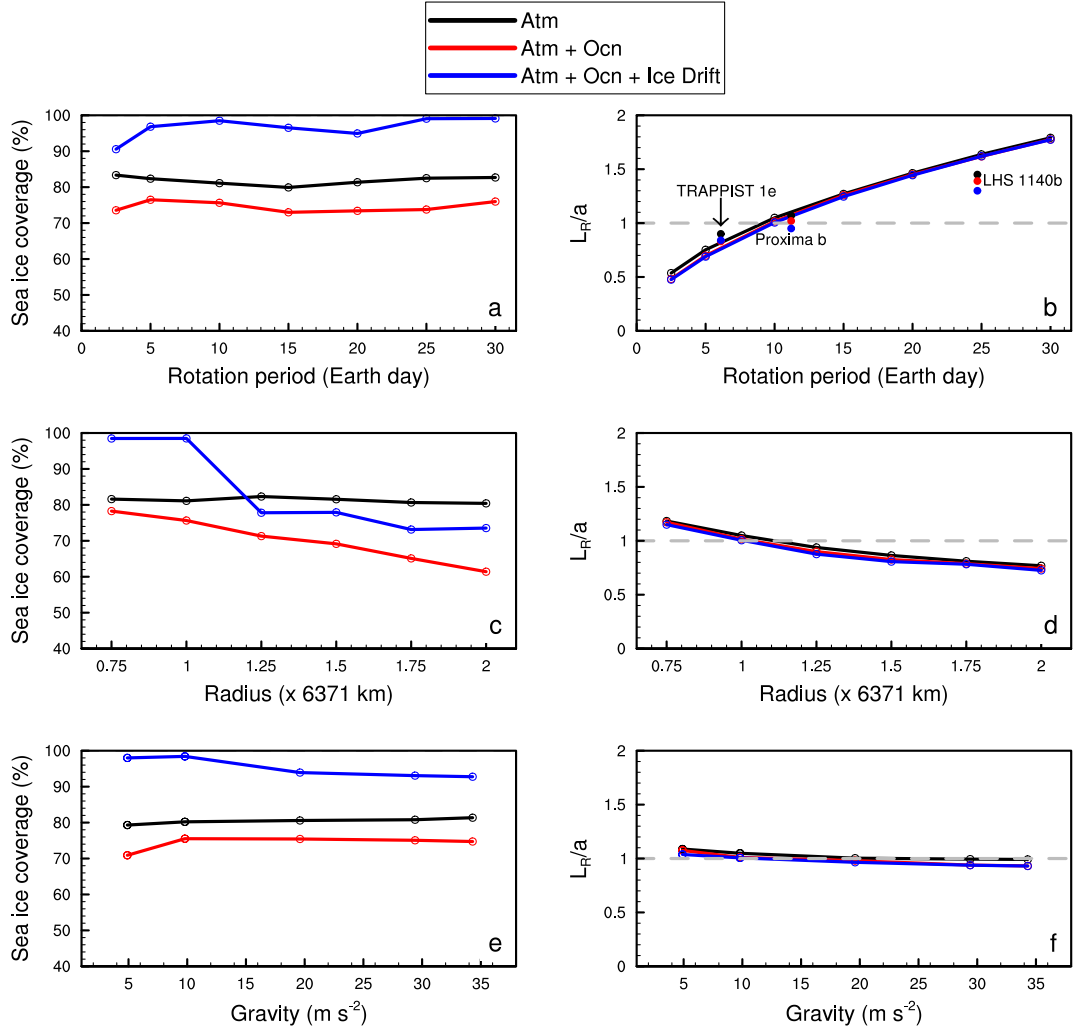

**Supplementary Figure 4**. Effects for varying planetary rotation period, radius, and gravity on the results. **a & b**, rotation period from 2.5 to 30 days; **c & d**, radius from 0.75 to 2.0 times Earth's value (6371 km); and **e & f**, gravity from 4.9 to 34.3 m s$^{-2}$. Left column: global-mean sea ice coverage, and right column: the ratio of equatorial Rossby deformation radius ($L_R$) to planetary radius ($a$). Black line: atmosphere-only run, red: coupled atmosphere-ocean run, and blue: fully coupled atmosphere-ocean-sea-ice run. The values of $L_R/a$ for the control runs of Proxima b, TRAPPIST-1e, and LHS 1140b are also shown in **b**. Be default, the stellar flux is 800 W m$^{-2}$, rotation period is 10 days, planetary radius is 1.0 Earth's value, gravity is 9.8 m s$^{-2}$, column air mass is $1.0 \times 10^4$ kg m$^{-2}$, and column $CO_2$ mass is ≈4.5 kg m$^{-2}$ in these experiments.

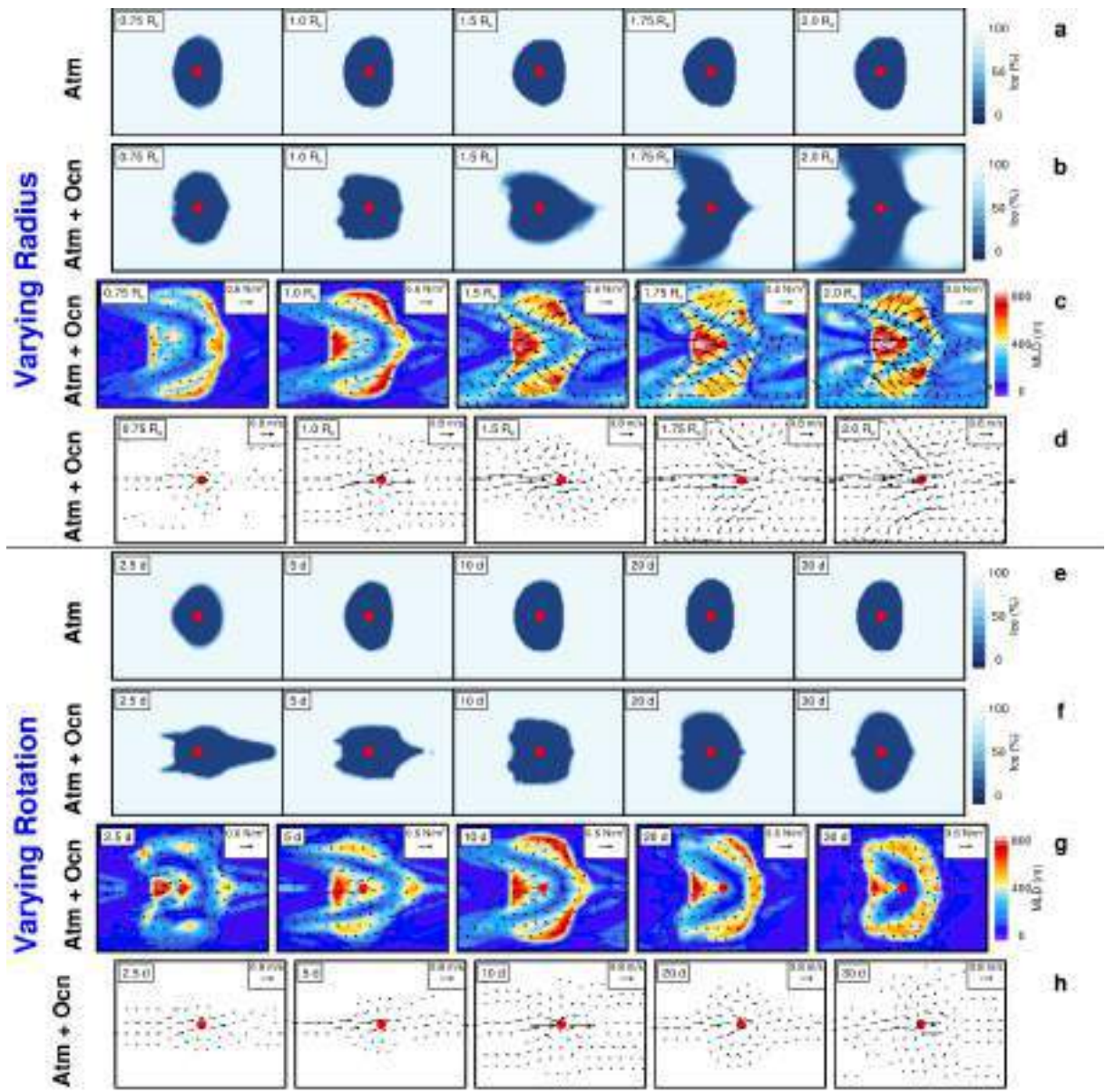

**Supplementary Figure 5**. Comparing the effects of between varying planetary radius (**a-d**) and varying rotation period (**e-h**). **a-d**, Sea ice concentration in the atmosphere-only experiments (**a**), and sea ice concentration (**b**), surface wind stresses over the ocean (vectors) and oceanic mixed layer depth (color shading) (**c**), and surface ocean currents (**d**) in the coupled atmosphere-ocean experiments. From left to right, the planetary radius is 0.75, 1.0, 1.5, 1.75, and 2.0 times Earth's value, respectively. **e-h**, Sea ice concentration in the atmosphere-only experiments (**e**), and sea ice concentration (**f**), surface wind stresses over the ocean (vectors) and oceanic mixed layer depth (color shading) (**g**), and surface ocean currents (**h**) in the coupled atmosphere-ocean experiments. From left to right, the rotation period is 2.5, 5, 10, 20, and 30 Earth days, respectively. By default, the radius is equal to Earth's value and the rotation period is 10 days. The x-axis is longitude from $0^{o}$ to $360^{o}$ and the y-axis is latitude from $90^{o}$S to $90^{o}$N in all the panels. All the variables in this figure are for equilibrium state except the panels **c** and **g**, for which annual-mean values of the second year during the integration are shown in order to avoid the strong effect of sea-ice melting and formation on the depth of the mixed layer.

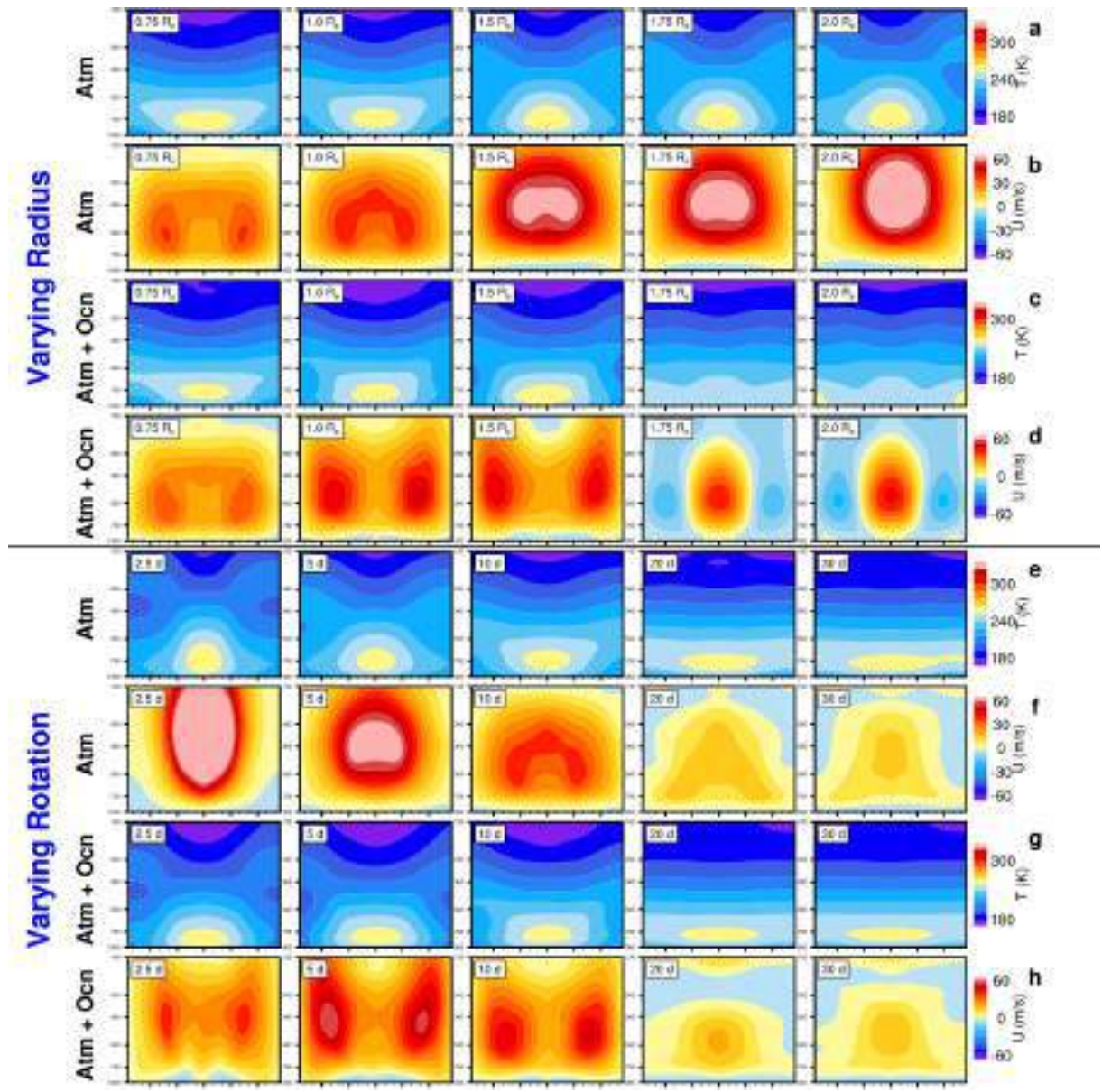

**Supplementary Figure 6**. Same as the Supplementary Figure 5 but for zonal-mean atmospheric temperature (**a**, **c**, **e**, & **g**) and zonal-mean zonal winds (**b**, **d**, **f**, & **h**). **a** & **b** are for atmosphere-only experiments of varying radius, and **c** & **d** are for coupled atmosphere-ocean experiments. **e** & **f** are atmosphere-only experiments of varying rotation, and **g** & **h** are for coupled atmosphere-ocean experiments. The x-axis is latitude from 90°S to 90°N and the y-axis is pressure from 1000 to 100 hPa in all the panels. By default, the radius is equal to Earth's value and the rotation period is 10 days. In the atmosphere-only runs, the effect of increasing planetary radius is similar to that of decreasing rotation period, but this equivalence breaks in the coupled atmosphere-ocean experiments because of their different effects on oceanic circulation and the following atmosphere-ocean interactions.

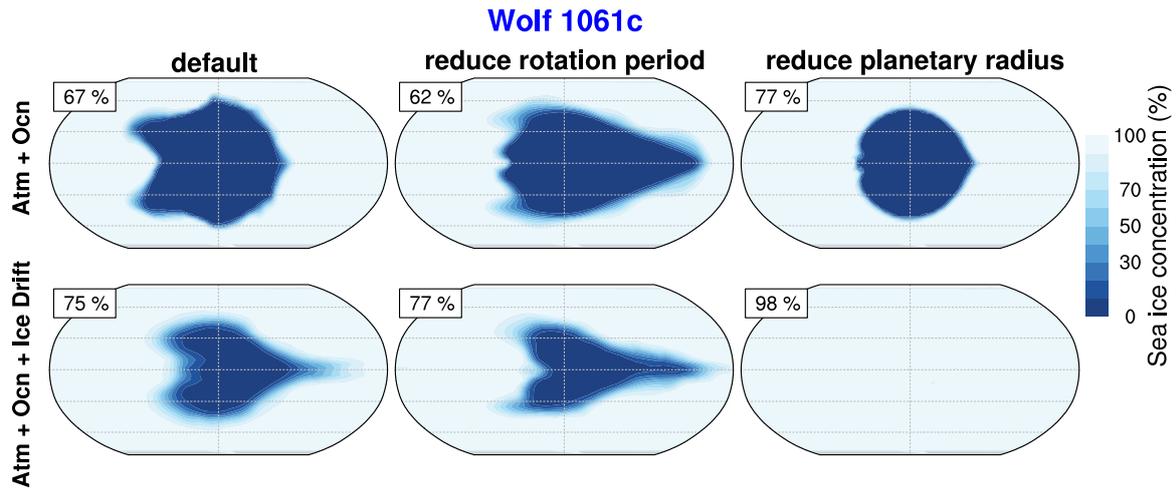

**Supplementary Figure 7**. The effects of varying rotation period and planetary radius on the sea ice concentration of Wolf 1061c under 1:1 orbits. Left column: the experiments with default parameters (same as those shown in Fig. 1 for clear comparisons). Middle column: decreasing the rotation period only from 17.9 days to 6.1 days (TRAPPIST-1e's value). Right column: decreasing the radius only from $10.2 \times 10^6$ km to $5.8 \times 10^6$ km (TRAPPIST-1e's value). The number in the upper left corner of each panel is the global-mean ice coverage. Varying the radius has a significant effect on the global-mean sea-ice coverage while varying rotation period has a much smaller effect. In all these experiments, sea-ice drift acts to increase the ice coverage.

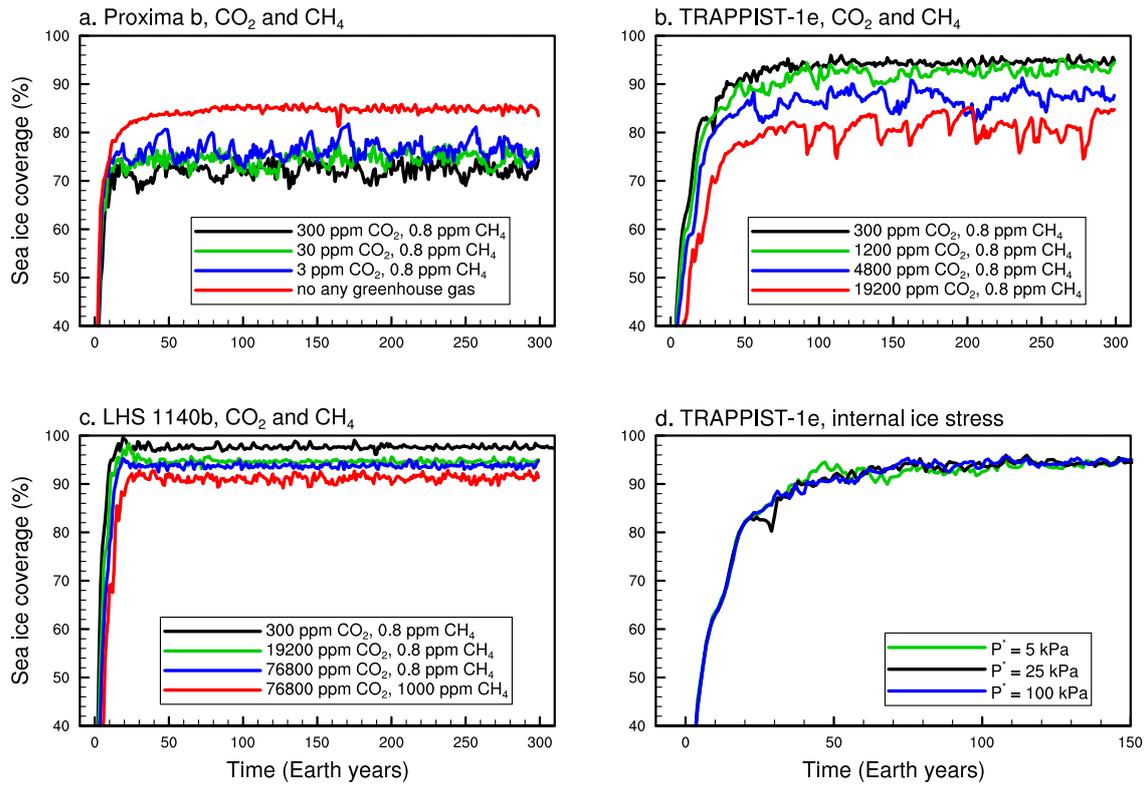

**Supplementary Figure 8**. Effects of greenhouse gas concentration and internal ice stress on the sea ice coverage in the fully coupled atmosphere-ocean-sea-ice experiments. **a, b,** & **c**, the effects of varying $CO_2$ and $CH_4$ concentrations on Proxima b, TRAPPIST-1e, and LHS 1140b, respectively. **d**, the effect of internal ice compressive strength ($P^*$) on TRAPPIST-1e. $P^*$ has a typical value of 25 kPa with a range of from 5 to 100 kPa in the different ocean basins of Earth[12]. The column air mass is $\approx 1.0 \times 10^4$ kg m$^{-2}$ in all these experiments.

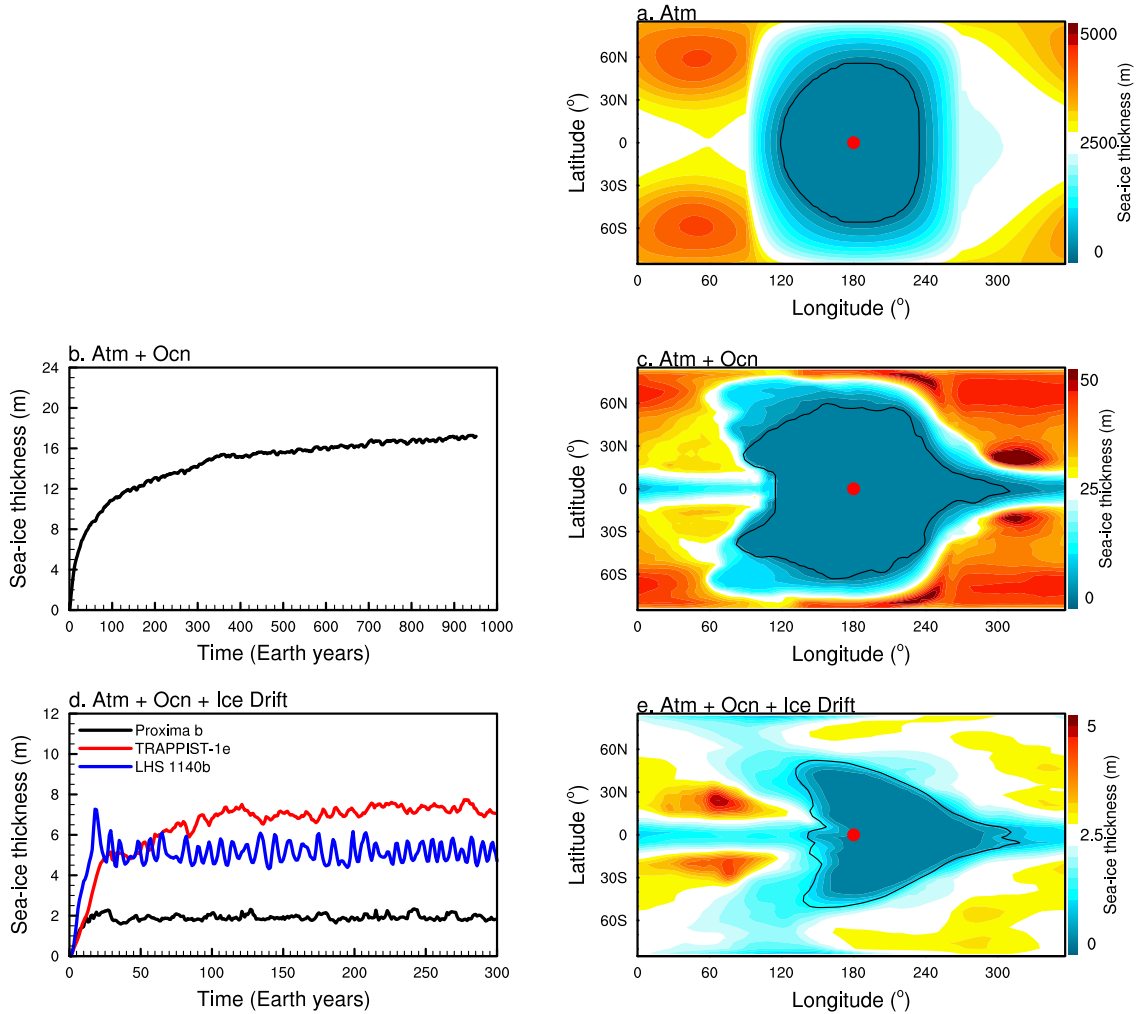

**Supplementary Figure 9**. Ice thicknesses. **a,** equilibrium sea ice thickness in the atmosphere-only experiment of Proxima b, calculated using the method of Menou (2013)[65]: $\kappa \Delta T / F_g$ where $\kappa$ is the thermal conductivity of ice (2.5 W m$^{-1}$ K$^{-1}$), $\Delta T$ is the temperature difference between the surface air and the seawater beneath the ice, and $F_g$ is the geothermal heat flux with a value of 0.06 W m$^{-2}$. **b-c,** time series of global-mean sea ice thickness (**b**) and equilibrium ice thickness (**c**) in the coupled atmosphere-ocean run of Proxima b. **d-e,** time series of global-mean sea ice thickness in the three fully coupled atmosphere-ocean-sea-ice control runs of Proxima b, TRAPPIST-1e, and LHS 1140b (**d**) and equilibrium sea thickness on Proxima b (**e**). The thin lines in **a, c,** & **e** are the sea ice edges with an ice concentration of 50%. Note the different scales for the five panels. Both oceanic heat transport and sea ice dynamics act to limit the sea ice thickness, leading to thin ice[20].

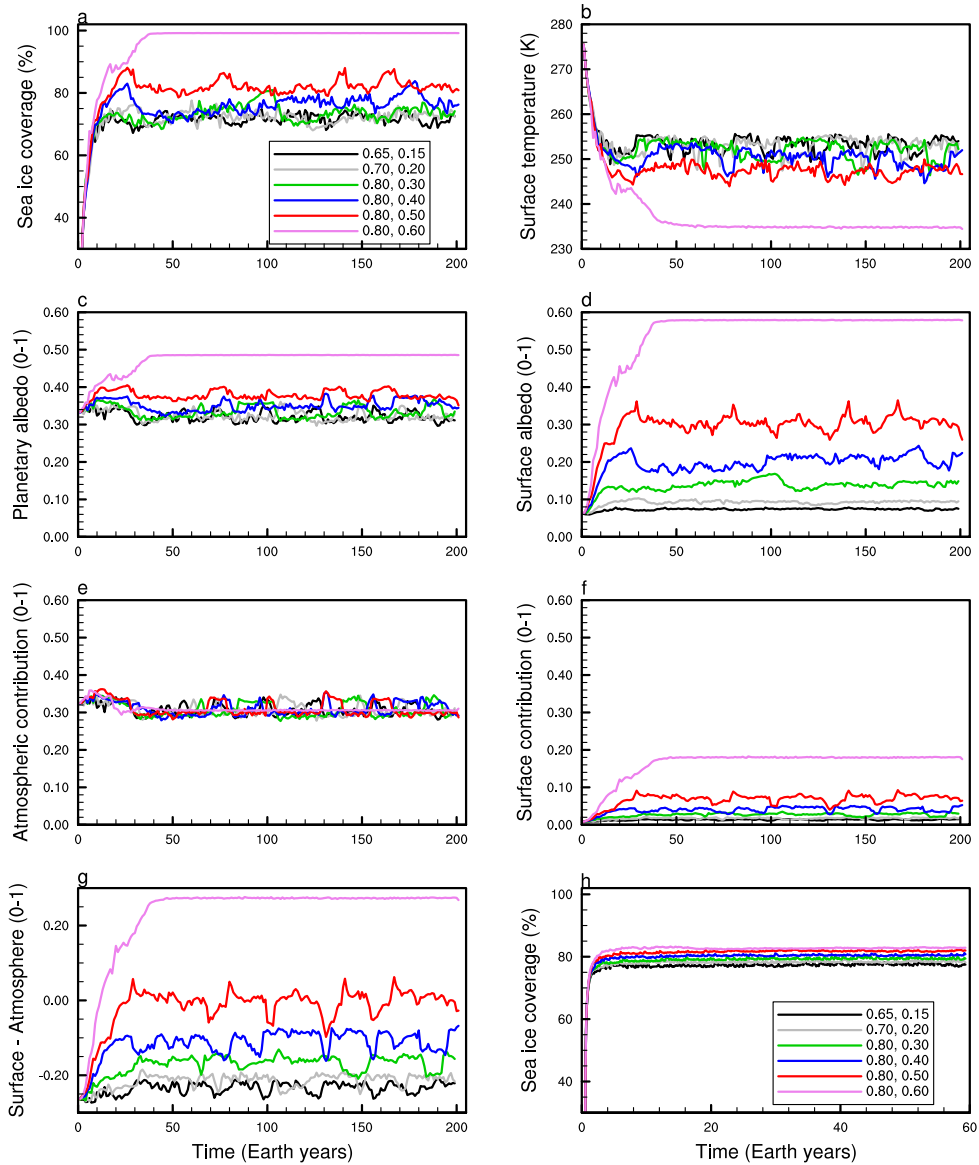

**Supplementary Figure 10**. Effects of sea ice albedo on the climate of Proxima b. Time-series of global-mean sea ice coverage (**a**), surface air temperature (**b**), planetary albedo (**c**), surface albedo (**d**), atmospheric contribution to the planetary albedo (= atmospheric reflection, **e**), surface contribution to the planetary albedo (**f**), surface albedo minus atmospheric reflection (**g**), and sea ice coverage (**h**). **a-g**, results of fully coupled atmosphere-ocean-sea-ice experiments, and **h**, results of atmosphere-only experiments. Six groups of different ice albedos of (0.65, 0.15) (the first number is for the visible band, and the second is for the near-infrared band), (0.7, 0.2), (0.8, 0.3), (0.8, 0.4), (0.8, 0.5), and (0.8, 0.6) were examined. The column air mass is $1.0 \times 10^4$ kg m$^{-2}$ and column $CO_2$ mass is ≈4.5 kg m$^{-2}$ in all these experiments. The surface climate is not sensitive to the ice albedo as long as it is smaller than the atmospheric reflection.

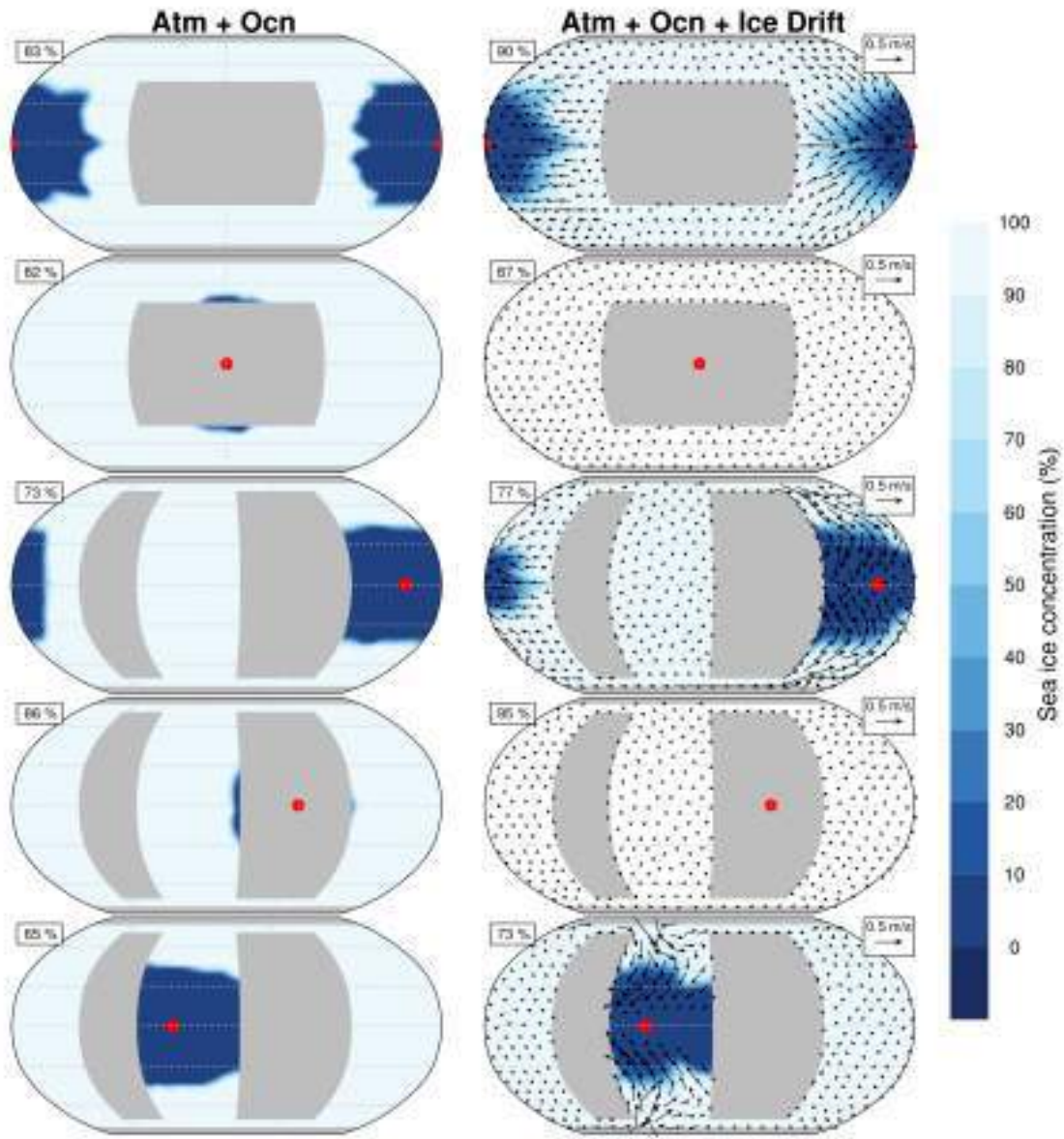

**Supplementary Figure 11. Effects of continents on the sea ice concentrations.** This figure is the same as Fig. 3 in the main text but for two idealized continents. The substellar point (red dot) locates at the center of the open ocean, at the center of the continent, over the relatively larger open ocean, at the center of the relatively larger continent, and over the relatively smaller open ocean, respectively. The ocean depth is 4000 m in all these experiments.

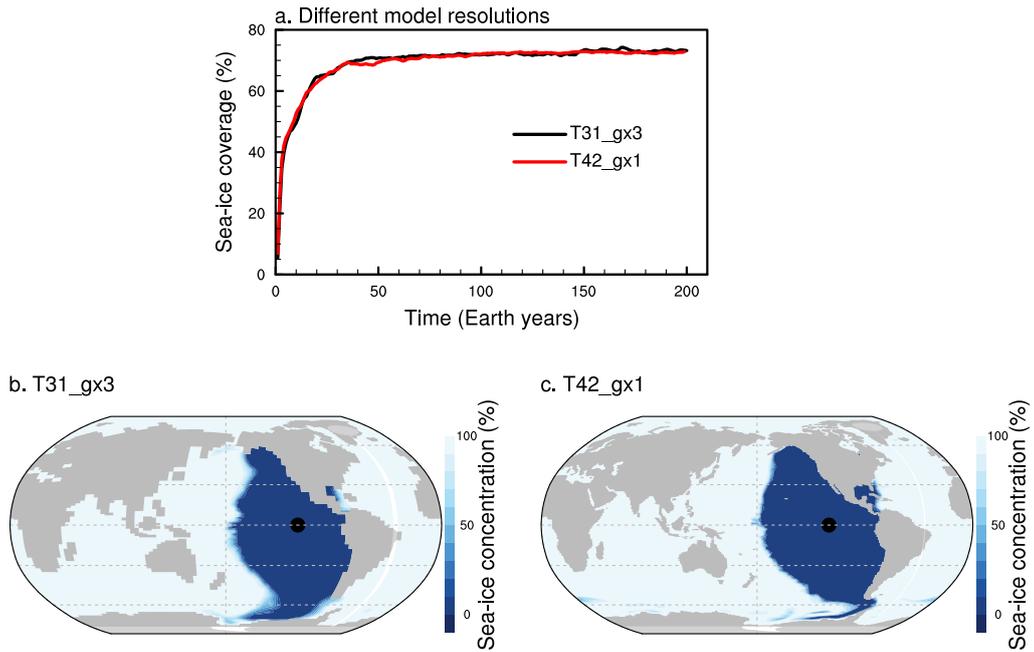

**Supplementary Figure 12**. The small effect of varying model resolution on the results in coupled atmosphere-ocean experiments under a 1:1 orbit. **a**, time series of global-mean sea-ice coverage in the experiments of two different resolutions, T31_gx3 and T41_gx1. **b**, equilibrium sea ice fraction in the T31_gx3 experiment. **c**, same as **b** but for the T42_gx1 experiment. The black dot is the substellar point, and Earth's continents and oceans were employed in the simulations. The parameters of TRAPPIST-1e were employed, column air mass is $1.0 \times 10^4$ kg m$^{-2}$, and column $CO_2$ mass is ≈4.5 kg m$^{-2}$.

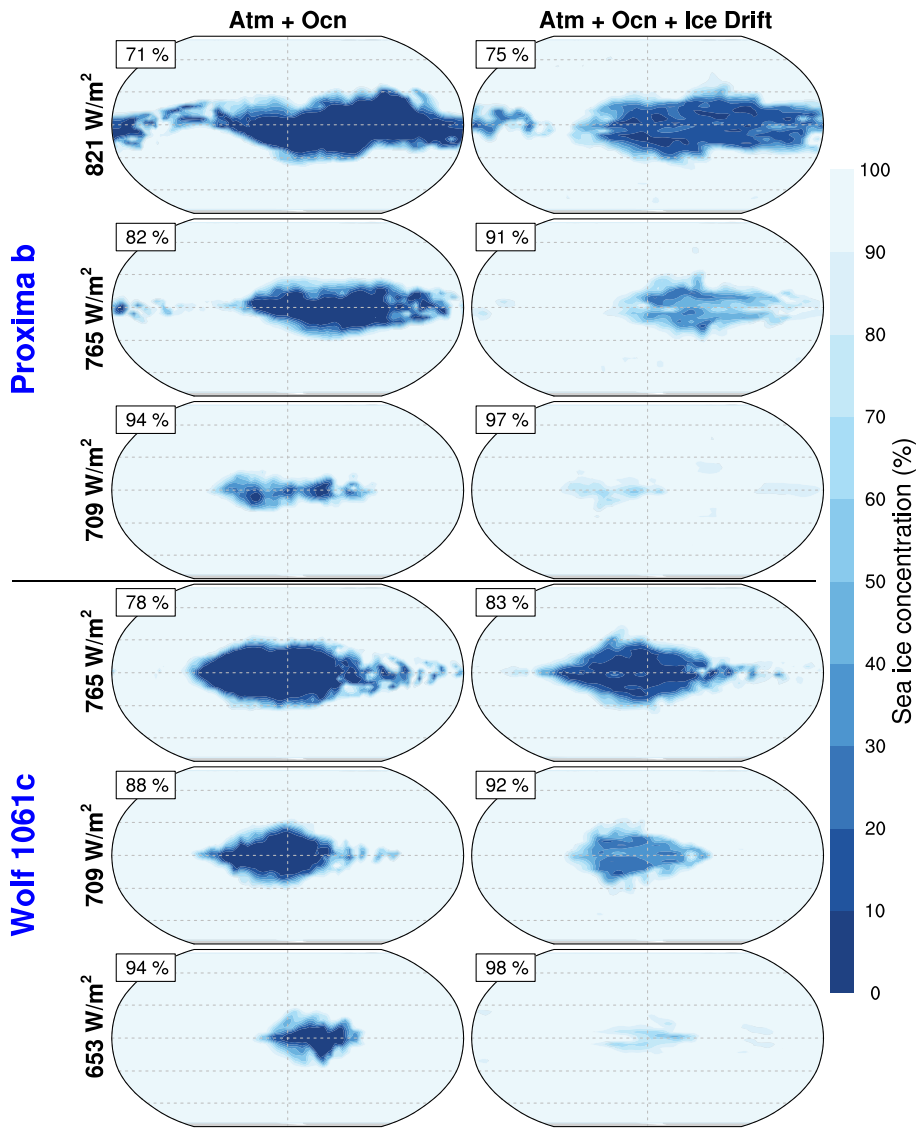

**Supplementary Figure 13**. Effects of sea-ice drift on the sea ice concentration (snapshot) for different stellar fluxes under 3:2 resonance orbits. The upper three rows: using the parameters of Proxima b but reducing the stellar flux from 887 to 821, 765, or 709 W m$^{-2}$. The lower three rows: using the parameters of Wolf 1061c but reducing the stellar flux from 819 to 765, 709, or 653 W m$^{-2}$. Left column: without sea-ice drift. Right column: with sea ice drift. The number in the upper left corner of each panel is the global-mean ice coverage. Sea-ice drift increases the ice coverage in all these experiments.

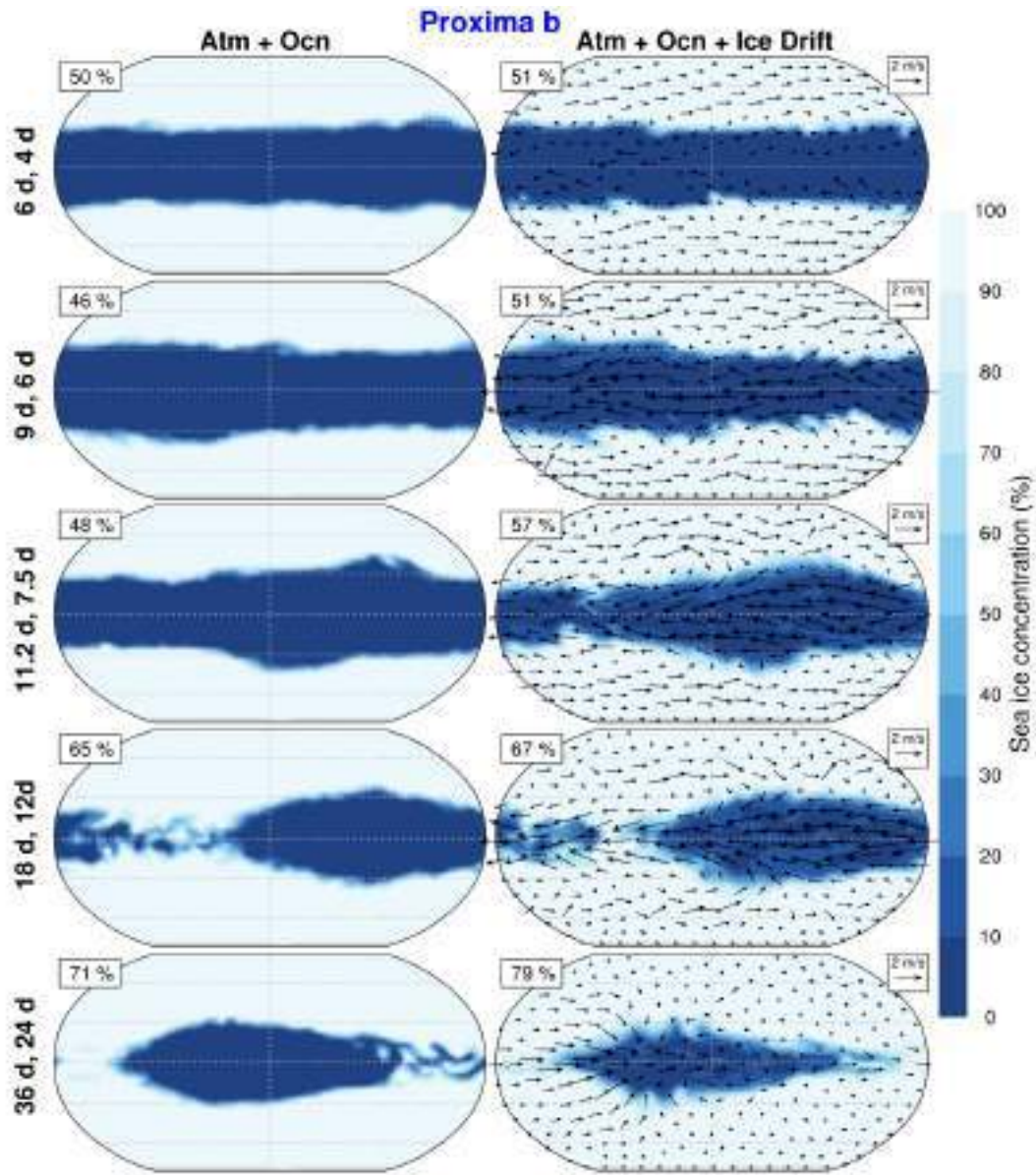

**Supplementary Figure 14**. Effects of varying rotation period on the sea ice concentration (snapshot) of Proxima b under 3:2 resonance orbits. From top to bottom, the orbital periods are 6, 9, 11.2, 18, and 36 days, and the corresponding rotation periods are 4, 6, 7.5, 12, and 24 days, respectively. Left column: without sea-ice drift. Right column: with sea ice drift. The vectors are sea ice velocities. The number in the upper left corner of each panel is the global-mean ice coverage. The center rows are the same as those shown in Fig. 4 for clear comparisons. Sea-ice drift increases the ice coverage in all these experiments.

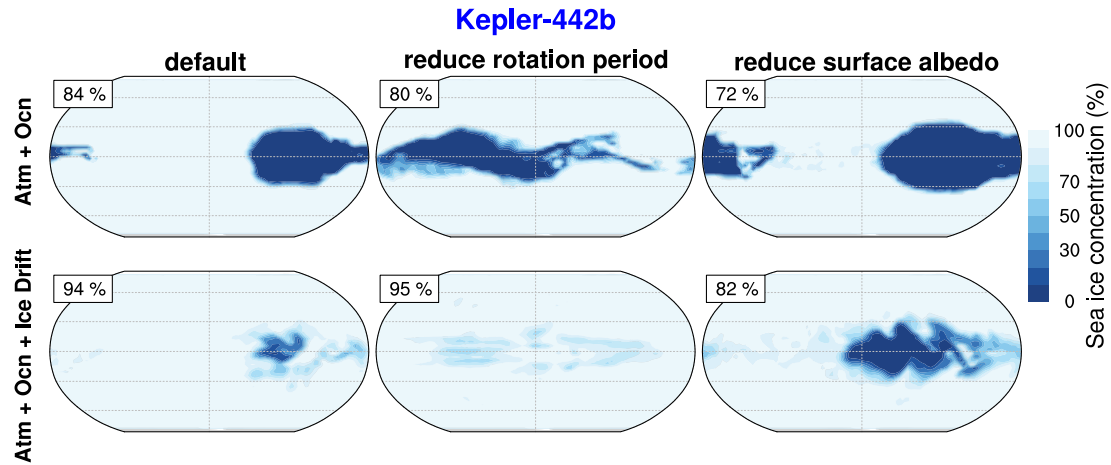

**Supplementary Figure 15**. Effects of varying rotation period and surface albedo on the sea ice concentration (snapshot) of Kepler-442b under 3:2 resonance orbits. Left column: the experiments with default parameters (same as those shown in Fig. 4 for clear comparisons). Middle column: decreasing the rotation period only from 112 days (default) to 11.2 days (Proxima b's value). Right column: decreasing the ice albedos only from (0.75, 0.30) to (0.65, 0.15) (see Supplementary Table 1). The number in the upper left corner of each panel is the global-mean ice coverage. The sea-ice coverage decreases significantly as decreasing the ice albedos.

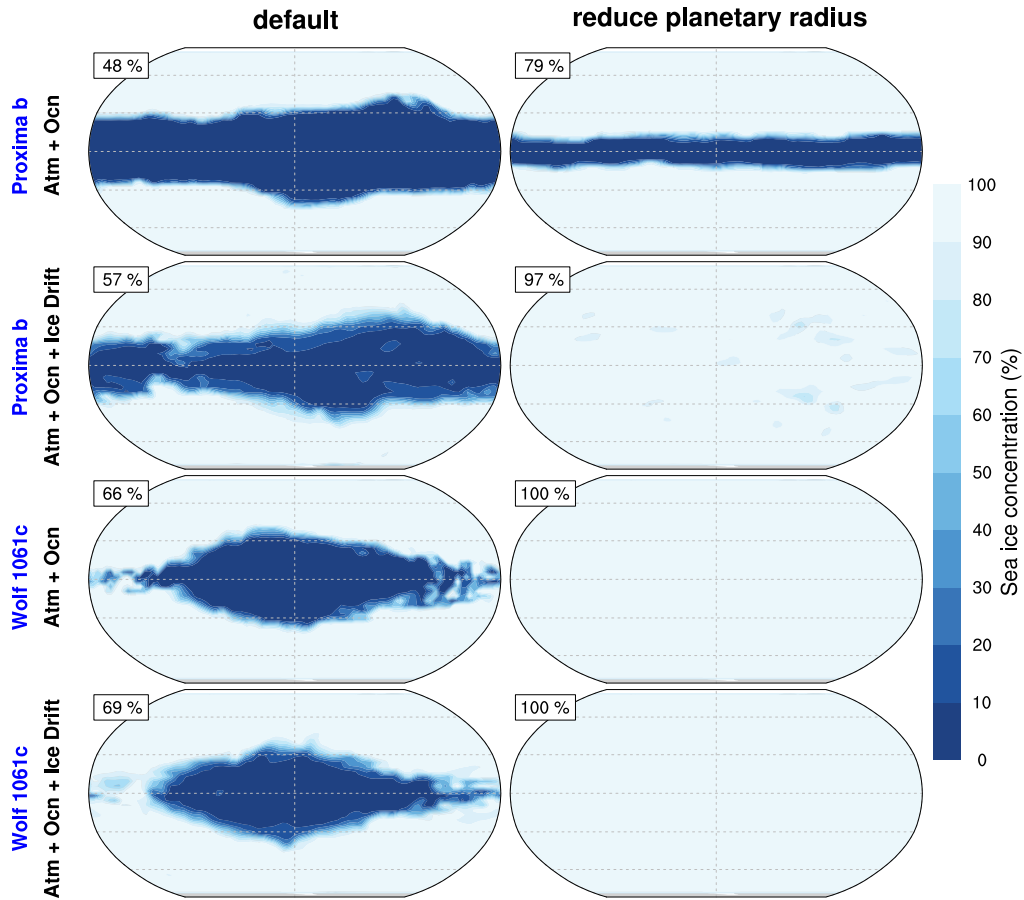

**Supplementary Figure 16**. Effects of reducing planetary radius on the sea ice concentration (snapshot) of Proxima b and Wolf 1061c under 3:2 resonance orbits. Left column: the experiments with default parameters (same as those shown in Figure 4 for clear comparisons). Right column: Decreasing the radius only from $8.1 \times 10^6$ km to $5.8 \times 10^6$ km (TRAPPIST-1e's value) for Proxima b and from $10.2 \times 10^6$ km to $5.8 \times 10^6$ km for Wolf 1061c. The number in the upper left corner of each panel is the global-mean ice coverage. The sea-ice coverage increases significantly as decreasing the radius.

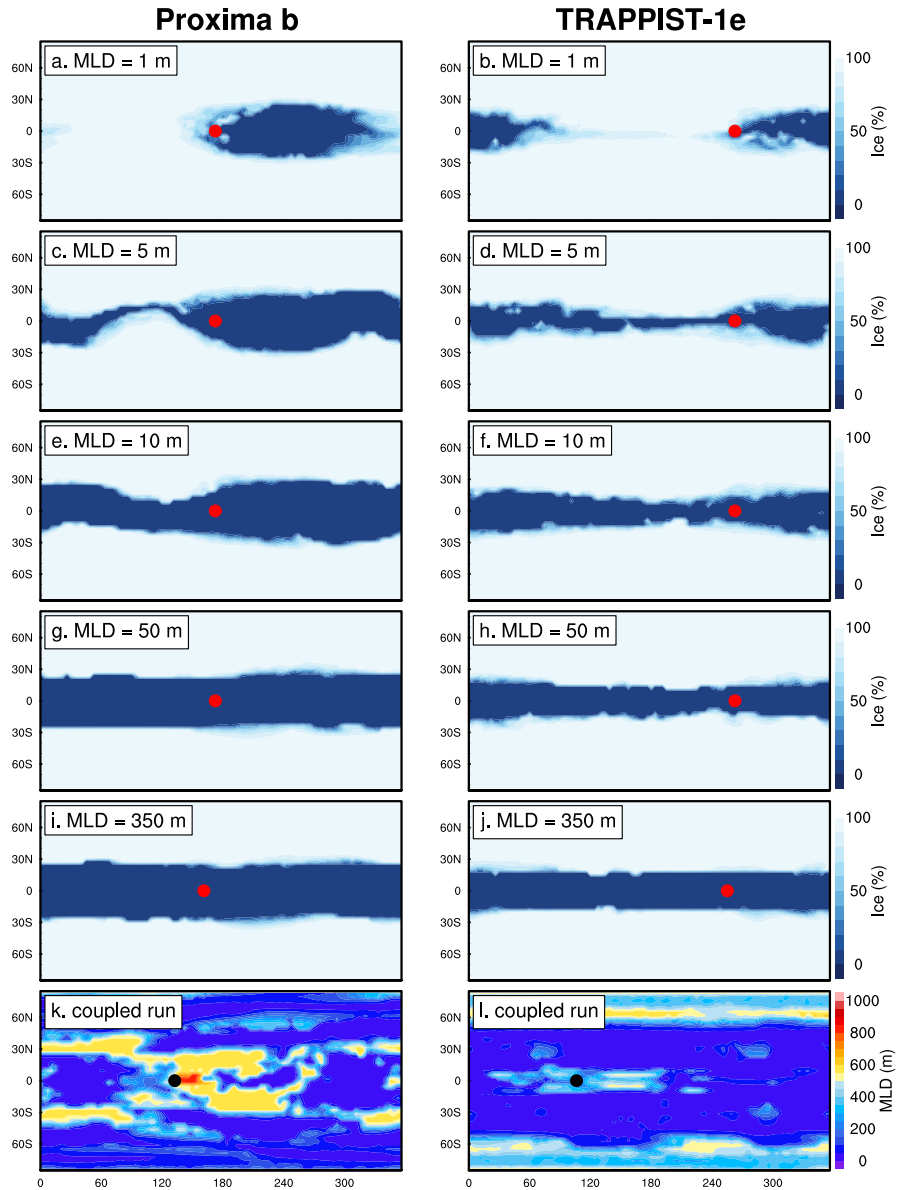

**Supplementary Figure 17**. Effect of ocean mixed layer depth (MLD) on the surface climate of 3:2 tidally locked orbits. Left column is for Proxima b, and right for TRAPPIST-1e. **a-j**, snapshots of sea ice concentration (%) for different MLDs of 1, 5, 10, 50, and 350 m (uniform), respectively, in the atmosphere-only simulations. **k-l**, snapshots of MLDs obtained in the coupled atmosphere-ocean experiments. The red dot in **a-j** and the black dot in **k-l** are the transient locations of the substellar points, which moves from east to west. The column air mass is $1.0 \times 10^4$ kg m$^{-2}$ and column $CO_2$ mass is ≈4.5 kg m$^{-2}$ in all these experiments. The sea ice coverage decreases as increasing mixed layer depth in the atmosphere-only experiments.

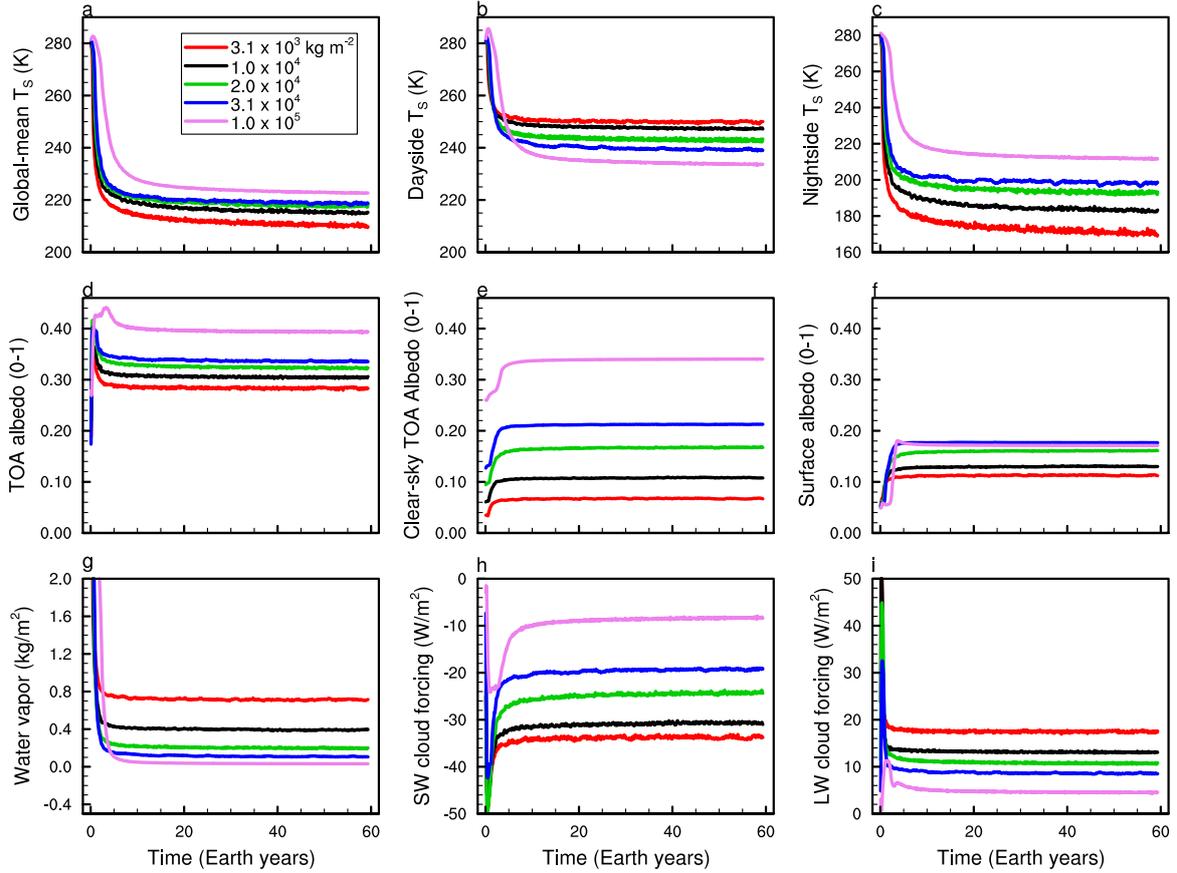

**Supplementary Figure 18**. Effects of background air mass ($N_2$) on the climate of LHS 1140b in the atmosphere-only experiments. Time series of global-mean surface air temperature ($T_S$, **a**), dayside mean $T_S$ (**b**), nightside mean $T_S$ (**c**), global-mean planetary albedo at the top of the atmosphere (TOA, **d**), clear-sky planetary albedo (**e**), surface albedo (**f**), vertically integrated water vapor amount (**g**), shortwave cloud forcing (**h**), and longwave cloud forcing (**i**). The column air masses are $3.1 \times 10^3$, $1.0 \times 10^4$, $2.0 \times 10^4$, $3.1 \times 10^4$, and $1.0 \times 10^5$ kg m$^{-2}$, respectively. The column $CO_2$ mass is $\approx 4.5$ kg m$^{-2}$ in all these experiments. Increasing background air mass promotes the onset of a snowball state.

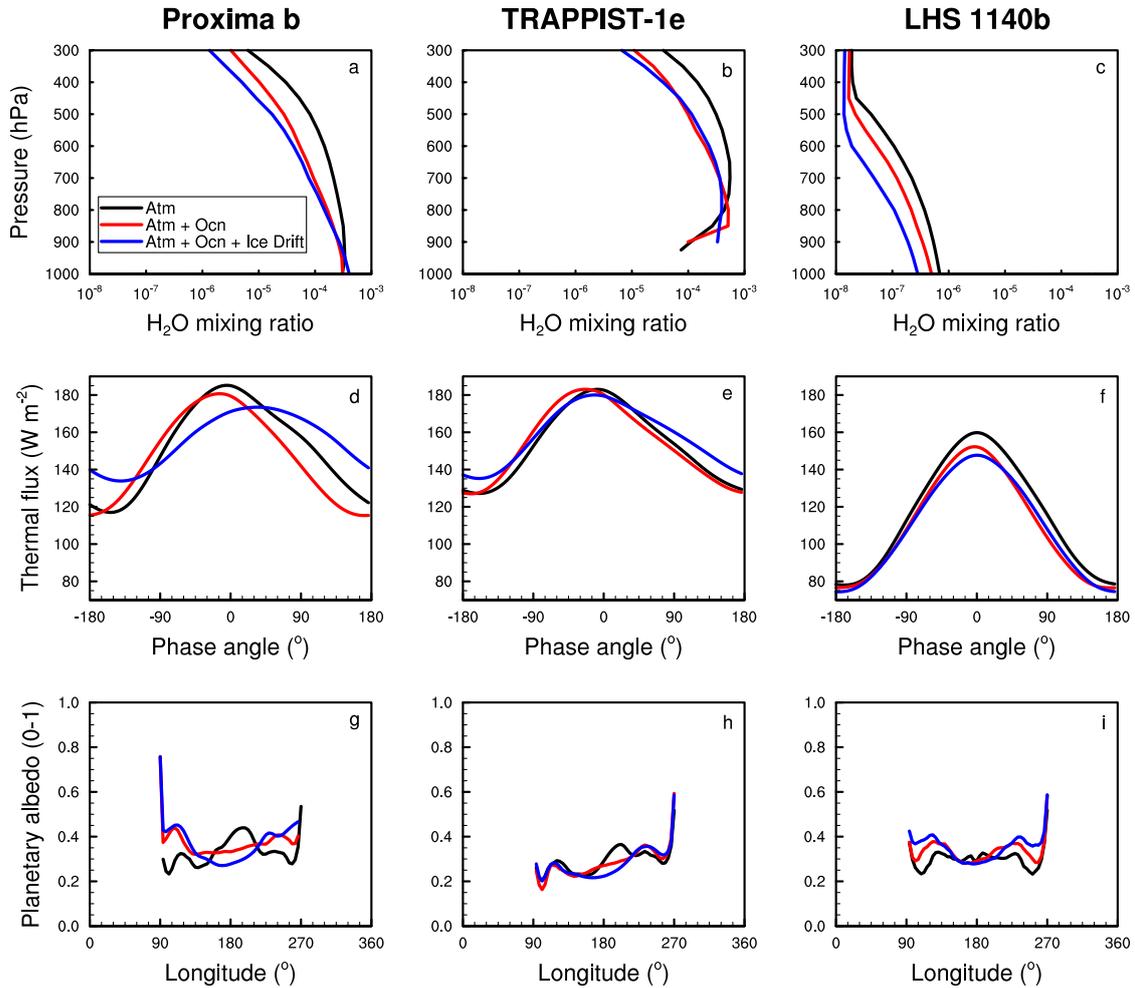

**Supplementary Figure 19**. Observational characteristics of the three planets in the control experiments. **a-c**, global-mean water vapor profiles at the terminators (average of the west terminator and the east terminator). **d-f**, thermal emission phase curve as a function of phase angle. The observer views the dayside of the planet at a phase angle of $0^o$ and the nightside at phase angles of $\pm 180^o$. **g-i**, meridional (north-south) mean planetary albedo as a function of longitude. Note that there is no albedo definition on the permanent nightside and the high albedo at the terminators is due to the large solar zenith angle of $\approx 90^o$. The differences in water vapor concentration, thermal emission flux, and planetary albedo between the three states, eyeball, lobster, and snowball, are small, and it would be hard to distinguish them in observations using recent telescopes.

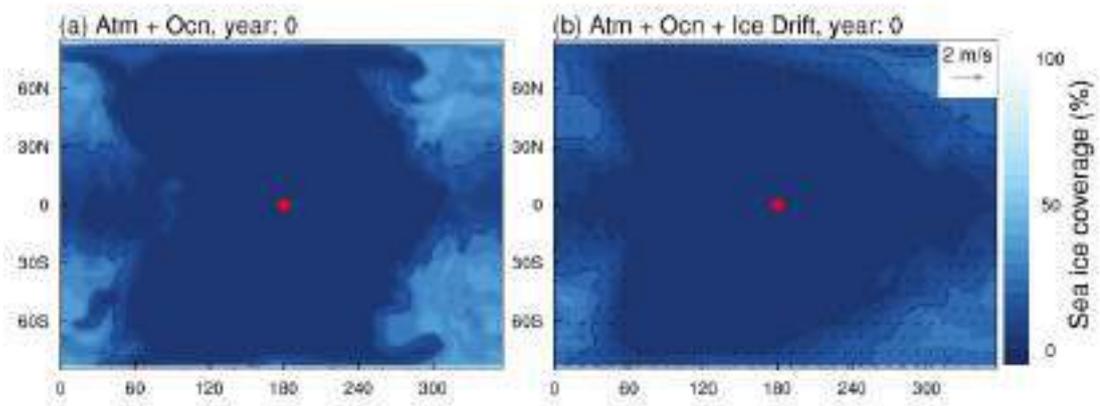

**Supplementary Video 1**. Evolution of sea ice concentration in the experiments of LHS 1140b. Left: coupled atmosphere-ocean run without sea-ice drift, and right: fully coupled atmosphere-ocean-sea-ice run with sea-ice drift. Color shading: sea ice concentration. Vector: sea ice velocity. Please watch this video online.